%% file: paper.tex
\documentclass{article}%
\usepackage[utf8]{inputenc}%
\usepackage{csquotes}%
\usepackage{authblk}
\usepackage{url}%
\usepackage{geometry}%
\usepackage{booktabs}%
\usepackage{color}%
\usepackage{soul}%
\usepackage[normalem]{ulem}%
\usepackage{comment}%
\usepackage{setspace}%
\usepackage{amsmath}
\usepackage[inline]{enumitem}%
\usepackage{soul}%
\usepackage{rotating}%
\usepackage{enumitem}%
\usepackage{tabularx}%
\usepackage{tcolorbox}%
\usepackage{color}%
\newif\ifshowrewrite
\showrewritefalse 
\usepackage{caption}%
\usepackage{subcaption}%
\usepackage{footmisc}

\usepackage[natbibapa]{apacite}
\usepackage[hidelinks]{hyperref} 
\hypersetup{
    colorlinks=false,
    linkcolor=black,
    citecolor=black,
    urlcolor=black,
}
\begin{document}%
%
%
%
\title{Mapping the Challenges of HCI:
An Application and Evaluation of ChatGPT for
Mining Insights at Scale
}%
%
%

\author[1]{Jonas Oppenlaender} 
\author[2]{Joonas Hämäläinen}
\affil[1]{University of Oulu, Center for Applied Computing, Faculty of Information Technology and Electrical Engineering, Oulu, Finland, \texttt{jonas.oppenlaender@oulu.fi}}
\affil[2]{University of Jyv\"askyl\"a, Software and Communications Engineering, Faculty of Information Technology, Jyv\"askyl\"a, Finland, \texttt{joonas.k.hamalainen@jyu.fi}}





\date{}




\maketitle%
\ifshowrewrite\color{orange}\fi

{%
\centering%
Corresponding author information: Jonas Oppenlaender
\\
Author e-mail:
\texttt{jonas.oppenlaender@oulu.fi}
\\
University of Oulu, Center for Applied Computing, Faculty of Information Technology and Electrical Engineering, Pentti Kaiteran katu 1, 90570 Oulu, Finland%
\\%
}%

\begin{abstract}%
{\noindent}
Large language models (LLMs) are increasingly used for analytical tasks, yet their effectiveness in real-world applications remains underexamined, partly due to the opacity of proprietary models. We evaluate ChatGPT (GPT-3.5 and GPT-4) on the practical task of extracting research challenges from a large scholarly corpus in Human-Computer Interaction (HCI). Using a two-step approach, we first apply GPT-3.5 to extract candidate challenges from the 879 papers in the 2023 ACM CHI Conference proceedings, then use GPT-4 to select the most relevant challenges per paper. This process yielded 4,392 research challenges across 113 topics, which we organized through topic modeling and present in an interactive visualization. We compare the identified challenges with previously established HCI grand challenges and the United Nations Sustainable Development Goals, finding both strong alignment in areas such as ethics and accessibility, and gaps in areas such as human-AI collaboration. A task-specific evaluation with human raters confirmed near-perfect agreement that the extracted statements represent plausible research challenges ($\kappa$ = 0.97). The two-step approach proved cost-effective at approximately US\$50 for the full corpus, suggesting that LLMs offer a practical means for qualitative text analysis at scale, particularly for prototyping research ideas and examining corpora from multiple analytical perspectives.
\end{abstract}%


\section{Introduction}%
\label{sec:introduction}%
%
Large Language Models (LLMs), such as ChatGPT \citep{chatgpt,gpt-4.pdf}, are increasingly used for analytical tasks across academic disciplines. A prominent application is their capacity to extract insights from text corpora and analyze them from multiple perspectives (e.g., \citet{3544548.3580907.pdf}).
This capacity for open-domain information retrieval, aimed at addressing specific research questions, builds on LLMs’ ability to follow instructions \citep{2109.01652.pdf,2203.02155.pdf,1706.03741.pdf,alpaca_farm_paper.pdf}
and perform in-context learning \citep{2212.10559.pdf,2005.14165.pdf,2205.11916.pdf,2209.11895.pdf}. LLMs may therefore support qualitative analyses of text corpora from diverse perspectives.


Yet concerns persist about the effectiveness of LLMs for such tasks. The opacity of state-of-the-art models, particularly the limited transparency surrounding training data for OpenAI’s LLMs \citep{gpt-4.pdf}, raises questions about data leakage from evaluation benchmarks \citep{2109.01652.pdf,2005.14165.pdf}
and the degree to which these models rely on memorization rather than genuine comprehension and reasoning \citep{2303.12712.pdf,2012.07805.pdf}. More broadly, LLMs are known to produce plausible but factually incorrect outputs (hallucinations) and may reflect biases present in their training data, both of which call for caution when these models are applied to research tasks.

LLMs are predominantly assessed using standardized benchmarks, with few exceptions (e.g., \citet{2303.12712.pdf}). While such benchmarks provide a reference point for evaluating model performance, overreliance on a limited set of benchmarks and standardized metrics is a growing concern \citep{2002.08512.pdf}. For researchers in academia and industry, this narrow focus on established evaluation frameworks may obscure opportunities to apply LLMs to real-world tasks \citep{215_reduced_reused_and_recycled_th.pdf,2012.05345.pdf,2303.12712.pdf}.


Extracting insights from an entire research field represents one such real-world task. Human-Computer Interaction (HCI) is a broad and fragmented discipline composed of diverse sub-communities \citep{10.1145/2556288.2556969}, and each year, thousands of researchers convene at the ACM Conference on Human Factors in Computing Systems (CHI), HCI’s flagship conference, to present work on a wide range of topics. In 2023, the CHI proceedings included 879 research papers across nearly 150 sessions. These proceedings provide a representative snapshot of current HCI research, yet the diversity of research activities and the volume of publications make it difficult to survey the field’s ongoing research challenges. We define a research challenge as an unresolved problem or question within a discipline. Identifying such challenges can provide insight into the current state of HCI and inform future research directions, a task that has grown in importance as the field expands and calls for synthesis work increase \citep{3544548.3581332.pdf}.

The scale and diversity of HCI make it difficult to systematically identify the field’s current research challenges. We therefore ask:
\begin{quote}%
    RQ1: \textit{How can large language models (LLMs) be used to automatically identify current research challenges in HCI?}
\end{quote}%

To investigate this question, we apply two LLMs in sequence, GPT-3.5 (version gpt-3.5-turbo-0301) and GPT-4 (version 4.0314), to extract insights from a large HCI text corpus. Because this corpus was published after both models were trained, it allows us to evaluate their capacity to reason over previously unseen data. Specifically, we ask:
\begin{quote}%
    RQ2: \textit{What are the current research challenges for HCI researchers?}
\end{quote}%

To explore this question, we analyze the CHI 2023 proceedings, which include 879 research papers selected for their breadth of coverage across current HCI research. Our analysis proceeds in two steps: first, GPT-3.5 extracts candidate research challenges from each paper, and then GPT-4 refines this output by selecting the five most relevant challenges per paper. GPT-3.5’s lower cost allows us to process a large volume of data within budget constraints, while GPT-4 provides stronger reasoning for the selection step.

Beyond identifying research challenges, we further ask:
\begin{quote}%
    RQ3: \textit{Do the current research challenges in HCI align with the field’s previously established ‘grand challenges’ and the United Nations’ Sustainable Development Goals (SDGs)?}
\end{quote}

HCI is a diverse and evolving field, with research expanding into areas that earlier grand challenges did not fully anticipate. While grand challenges were designed to provide long-term guidance, a disconnect between current research and these challenges may suggest the need for their revision. Similarly, if current HCI research does not align with the Sustainable Development Goals, this may point to gaps in the field's engagement with global concerns.



Our analysis identified 4,392 research challenges across 113 topics, from accessibility to visualization. These topics encompass both established HCI domains and emerging areas of inquiry. Intertopic distance analysis reveals well-connected research themes across HCI and highlights opportunities for future investigation. The extracted challenges are more specific and actionable than previous grand challenges, with some that address areas not anticipated by earlier frameworks. At the same time, we observe limited alignment with sustainability in certain domains, which raises questions about the field's engagement with global priorities. To evaluate the extraction approach, we assessed whether the extracted statements constitute research challenges and measured inter-rater agreement, which yielded a Cohen's $\kappa$ of 0.97. The two models complement each other for cost-efficient insight extraction, a result that suggests LLMs can support large-scale qualitative analysis from multiple perspectives. The full dataset of 4,392 HCI research challenges is publicly available for further analysis by the research community.

This paper makes two key contributions:
\begin{enumerate}
    \item
    \textbf{Identification of Research Challenges in HCI
    (Section \mbox{\ref{sec:results}}):}

We identify current research challenges in Human-Computer Interaction and present them through an interactive visualization that provides a structured overview of the field and its emerging opportunities.
We compare these findings with HCI’s grand challenges and the United Nations’ Sustainable Development Goals to assess the field’s alignment with broader global objectives.
The dataset (see OSF repository\footnote{\href{https://osf.io/fbsq7/?view_only=b6b86e89fe444a74a33950b452a55242}{https://osf.io/fbsq7/?view\_only=b6b86e89fe444a74a33950b452a55242}}) and the interactive visualization\footnote{see \href{https://HCI-research-challenges.github.io}{https://HCI-research-challenges.github.io}} are openly shared to support further research.

    \item
    \textbf{Evaluation of LLMs for Cost-Efficient Insight Mining
    (Section \mbox{\ref{sec:evaluation}})}:
We evaluate a two-step approach that uses GPT-3.5 and GPT-4 for extracting insights from a large text corpus. This method extends researchers’ capacity to conduct qualitative analysis at scale. We critically examine the strengths and limitations of this approach, with attention to its applicability for cost-efficient text mining.
\end{enumerate}

Evaluating LLMs on real-world tasks, beyond standardized benchmarks, is important for understanding the practical capabilities of GPT-3.5 and GPT-4. Our goal is not to compare these models directly but to assess how GPT-4 can complement GPT-3.5 for cost-effective insight extraction from text corpora. Working with LLMs involves both direct costs (e.g., API usage) and indirect costs (e.g., iterative prompt design, experimentation, and prototyping), and we prioritize cost-efficiency in this study, which is particularly valuable for pilot studies and for analysis of data from multiple perspectives. While our focus is on HCI, the approach and findings may inform similar applications in other academic disciplines, with appropriate domain-specific adaptations.

\section{Related Work}%
\label{sec:relatedwork}%

\subsection{Language Models}%
\label{sec:languagemodels}%

Large language models (LLMs) process sequences of tokens and generate probability distributions over potential next tokens, producing text by sampling from these distributions in response to prompts. Trained on large internet text corpora, LLMs are increasingly used as retrieval-based systems that function as a new form of search engine \citep{3544548.3580688.pdf}. This process is commonly referred to as knowledge extraction \citep{2012.07805.pdf} or knowledge retrieval \citep{3543873.3587655.pdf}.

Despite their utility, LLMs face several challenges that affect their reliability, including data biases, hallucinations, and limited interpretability \citep{IJHCI4}. For example, when queried for a scientific reference, an LLM may generate fabricated citations to non-existent articles, illustrating the risk of hallucinations \citep{3571730.pdf,2305.13534.pdf,2306.05426.pdf}.

Hallucinations occur when a model generates plausible but factually incorrect content, a tendency that is well documented in the literature \citep{3571730.pdf,2305.13534.pdf}. In academic research contexts, hallucinations may take specific forms: models can fabricate citations, attribute findings to the wrong sources, or produce confident summaries of relationships absent from the source material. The risk is particularly pronounced in zero-shot knowledge extraction, where no reference text is provided. Training data biases compound this problem, as LLMs may reproduce stereotypes, encode cultural assumptions, or reflect distributional properties of their training corpora rather than the content of a given text \citep{2005.14165.pdf,Bommasani2021FoundationModels,2021.acl-long.330.pdf}. These limitations make it important to distinguish between tasks that draw on an LLM's parametric knowledge and those that ground the model in a provided text, a distinction central to our study.

Beyond retrieval, LLMs can also perform reasoning tasks through in-context learning \citep{2212.10559.pdf,2005.14165.pdf,2205.11916.pdf,2209.11895.pdf}. In this mode, the model processes a given text and reasons over it without fine-tuning, which allows LLMs to address a range of downstream tasks that extend beyond factual recall.

This study explores the use of in-context learning for insight mining, with the goal of extracting answers to specific research questions from a text corpus. Because the same corpus can be analyzed from multiple perspectives, cost-efficiency is a central concern. The following section situates this task within related work on automated text analysis in scholarly literature.

\subsection{Insight Mining with LLMs}%
\label{sec:infoextraction}%
Our research does not map neatly onto a single established field. This section situates our approach by examining its relationship to existing tasks and the performance of OpenAI’s LLMs in these contexts.


\textit{Information extraction} (IE) involves identifying structured information, such as entities, events, and their relationships, from unstructured text \citep{ijcai16a.pdf}. Researchers have developed various open information extraction (Open-IE) systems over the years (e.g., \citet{clausie-www13.pdf,D12-1048.pdf,IJCAI07-429.pdf}), and ChatGPT has shown strong performance on Open-IE tasks despite a tendency toward overconfidence in its outputs \citep{2304.11633.pdf}, consistent with its effectiveness on a range of zero-shot natural language tasks \citep{2302.06476.pdf}. Unlike IE, which produces structured output, our approach yields unstructured textual insights.


\textit{Document summarization} aims to capture a document’s main concepts concisely while reducing redundancy \citep{1f51871036db972480390a52d1d11375.pdf}. Extractive summarization, in particular, selects key passages to represent a document’s core content \citep{3544548.3581260.pdf}. Our work differs from summarization in that we seek to identify research challenges within a corpus rather than to condense individual documents.


\textit{Question answering} (QA) involves producing responses to specific queries from textual input \citep{TREC8}. Closed-domain QA requires interpreting a query in relation to a given text, a task where LLMs have shown proficiency through multi-step reasoning \citep{2208.14271.pdf}. Traditional QA systems aim to answer a broad range of questions, whereas our research focuses on a single, in-depth question applied across an entire corpus. Although there is conceptual overlap, our task demands a level of analytical depth that extends beyond conventional QA.


\textit{Qualitative analysis} seeks to uncover patterns, themes, and underlying meanings within data \citep{braun2006.pdf,braun2019.pdf}. Grounded in established theoretical frameworks \citep{glaserstrauss}, qualitative methods are widely used in disciplines that require in-depth textual analysis, yet they are difficult to scale to large datasets \citep{deterding2018.pdf}.
%
Recent studies have explored the use of LLMs for qualitative analysis, with ChatGPT showing promise in assisting with qualitative coding \citep{2304.07366.pdf,3581754.3584136.pdf}. Our approach shares the interpretative goals of qualitative analysis but uses LLMs to analyze text corpora at scale.


\textit{Knowledge discovery} (KD) focuses on identifying patterns, trends, and insights within large datasets through synthesis and interpretation rather than simple retrieval \citep{240455.240464.pdf}. While our use of LLMs shares this objective, traditional KD methodologies rely on computational and statistical tools rather than language models. Additionally, knowledge discovery in databases (KDD) \citep{240455.240464.pdf} is closely associated with database-driven analysis, whereas our approach does not involve structured databases.


\textit{Information retrieval} (IR) is concerned with locating relevant information within large datasets in response to user queries. IR is effective for retrieving specific content but typically lacks interpretative depth, which makes it less suited to our goal of producing analytical insights from a text corpus.


\textit{Text mining} involves analyzing unstructured text to extract meaningful information and identify patterns, typically through statistical and algorithmic methods rather than language models. To differentiate our work from traditional text mining, we adopt the term \textit{insight mining}, which refers to the use of LLMs to extract analytical insights from large text corpora in response to specific research questions.



Across these knowledge-intensive tasks, maintaining faithfulness and reducing hallucinations remain central concerns \citep{2020.acl-main.173.pdf,2022.ecnlp-1.27.pdf,2208.14271.pdf,2301.00303.pdf}. ChatGPT has shown the ability to analyze text from multiple perspectives, supporting ``What if?'' analyses \citep{3544548.3580688.pdf}. Our research extends this capacity to an entire corpus of documents to extract targeted insights. The following section examines how LLMs are typically evaluated for such tasks and discusses the limitations of standard evaluation methods in real-world applications.

\subsection{Limitations of Large Language Models for Insight Mining}%
\label{sec:llmlimitations}%

When LLMs are applied to insight mining tasks, several limitations deserve attention. Because these models are trained on web-scale corpora, they may introduce systematic biases into extracted insights, with outputs that reflect the distributional properties of the training data rather than the content of the text being analyzed \citep{2021.acl-long.330.pdf,1707.09457.pdf}. In academic contexts, well-represented research areas may receive disproportionate attention, while underrepresented topics may be overlooked or distorted \citep{Bommasani2021FoundationModels}. LLMs have also been shown to encode latent opinions \citep{3544548.3581196.pdf} and cultural values \citep{2203.07785.pdf} in their outputs, which could affect how research challenges are framed or prioritized during extraction. LLMs may further exhibit sycophantic tendencies, producing outputs that align with perceived user expectations rather than the source material \citep{2212.09251.pdf,2310.13548.pdf}.

The risk of hallucination also differs depending on task design. In zero-shot knowledge extraction, where no source text is provided, hallucination rates tend to be higher \citep{3571730.pdf,2305.13534.pdf}. In-context reasoning tasks, where the model is given a text and asked to extract information from it, present a different risk profile: responses are grounded in the provided context, which reduces but does not eliminate the possibility of fabricated content. More subtle forms of hallucination may persist, as models may overgeneralize from specific claims, conflate findings across sections of a paper, or generate research challenges that sound plausible but are not directly supported by the source text \citep{2305.13534.pdf}. To address these risks, we employ a two-step extraction process paired with human evaluation (Section~\ref{sec:evaluation}) to mitigate the effect of ungrounded outputs on our results.
%
\subsection{Evaluating Large Language Models on Real-World Tasks}%
\label{sec:evaluatingllms}%

As large language models are increasingly integrated into applied settings, rigorous evaluation on real-world data and tasks becomes correspondingly important. Academic research has traditionally relied on task-specific benchmark datasets and performance metrics for this purpose (e.g., \citet{2304.11633.pdf}), and benchmarks remain the standard evaluation methodology in machine learning \citep{2303.12712.pdf}. Yet benchmark datasets and automated evaluation metrics \citep{BLANC,2022.naacl-main.153.pdf} may not fully capture a model’s effectiveness in practical applications. Benchmark datasets can also diverge from real-world use cases, which makes them susceptible to distribution shifts \citep{3544548.3581482.pdf}.

Despite these limitations, academic and industry research frequently relies on a small set of standardized evaluation frameworks \citep{215_reduced_reused_and_recycled_th.pdf,2012.05345.pdf,2303.12712.pdf}. There are exceptions: \citet{2303.12712.pdf} investigated GPT-4’s performance on specific edge cases, and \citet{2309.17421.pdf} curated qualitative samples for evaluating GPT-4V. Qualitative evaluations, however, remain uncommon in LLM research.

This predominant focus on benchmarks and standardized metrics risks overlooking opportunities to evaluate LLMs in real-world contexts \citep{215_reduced_reused_and_recycled_th.pdf,2012.05345.pdf,2303.12712.pdf}. Emerging applications in robotics \citep{ChatGPT___Robotics.pdf} and behavioral simulations \citep{2304.03442.pdf,2310.05418.pdf} suggest growing interest in applied evaluation, yet real-world task evaluation of LLMs remains underexplored \citep{3544548.3581503.pdf}.

In this study, we address the real-world task of identifying research challenges in Human-Computer Interaction. We compare our findings with established HCI grand challenges to examine how the field’s current research priorities relate to previously articulated goals.

\subsection{Grand Challenges in HCI}%
\label{sec:grand}%

Grand challenges are complex, high-impact problems that demand interdisciplinary collaboration and sustained effort over time. They are typically characterized by their global or societal significance, technical complexity, and the need for a coordinated research response. In HCI, grand challenges have historically served to direct attention, resources, and research activity toward pressing issues. We review two influential frameworks that have outlined such challenges for the field, as they provide context for our own findings.


\citet{p24-shneiderman.pdf} identified 16 grand challenges for HCI that span areas such as lifelong learning, resource conservation, and persuasive technologies. Their framework places particular emphasis on accountability and responsibility in healthcare-related HCI applications, and reflects the diversity of research priorities within the field.

\citet{10.1080@10447318.2019.1619259.pdf} proposed another set of HCI challenges centered on the interaction between humans and technology. Their framework addresses health, well-being, creativity, accessibility, ethics, privacy, security, and human-environment interaction, and reflects the expanding scope of HCI research and its growing relevance to society.


These grand challenges connect to the United Nations’ Sustainable Development Goals (SDGs), particularly in areas related to health, well-being, education, and inclusive societies \citep{unitednations2015}. The shared emphasis on sustainable technologies, ethics, accessibility, and secure information sharing suggests that HCI research priorities are shaped not only by technological and human-centered concerns but also by broader societal development goals.


In this study, we use these two sets of grand challenges as a reference framework to contextualize our findings and compare them with contemporary research challenges extracted from the CHI 2023 proceedings. We also examine the extent to which current HCI research priorities align with the UN’s SDGs. The following section details our data collection and processing methods.


\section{Method}%
\label{sec:method}%
Our approach consists of several phases: data acquisition and
preprocessing, challenge extraction with large language models, topic
modeling, and visualization (\autoref{fig:process}). The extraction
phase uses a two-step design in which we first identify candidate
challenges with GPT-3.5 and then refine them with GPT-4. We describe
each phase in the following subsections.%

\input{FIG-PROCESS}

%
%
\subsection{Data Acquisition
}%
\label{sec:information-acquisition}%
\label{sec:corpus}%
We manually downloaded all full research papers from the CHI 2023
proceedings on the ACM Digital Library (ACM--DL) and extracted text
from each paper with the textract Python library (version 1.6.5). For
one paper where text extraction failed, we first applied OCR software
to embed text into the PDF and then ran textract on the result.

\begin{figure}[!htb]%
\centering%
\includegraphics[width=.85\linewidth]{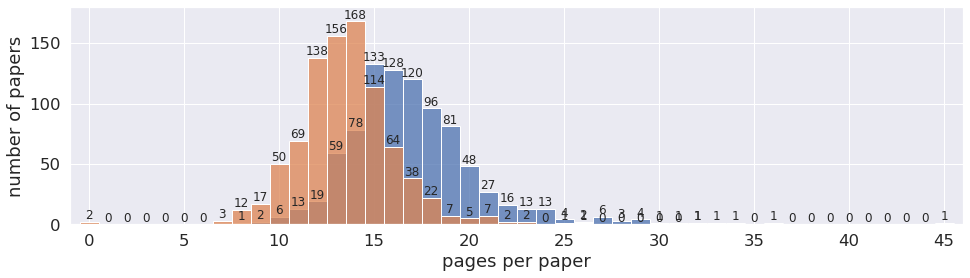}%
\caption{Histogram of pages per research paper in the CHI 2023 proceedings. Blue: total number of pages per paper, including references and appendices. Orange: approximate number of content pages per paper.}%
\label{fig:proceedings}%
\end{figure}

The CHI 2023 proceedings comprise 879~research papers formatted in
ACM's double-column template. Paper lengths range from 8 to 45 pages
when references and optional appendices are included. To better
approximate the number of content pages, we counted pages up to the
final occurrence of the word ``References'' in each document, which
excludes reference lists and appendices (\autoref{fig:proceedings}).
Under this measure, content pages range from 7 to 32 ($M = 13.5$,
$SD = 2.6$).

The papers at CHI 2023 were presented in 148~diverse thematic sessions,
some with overlapping topics. For instance, mental health appeared in
three dedicated sessions with six papers each. This volume and
thematic diversity make it difficult to gain an overview of the
field's current research challenges, which motivates the extraction
approach we describe in the following subsections.

\subsection{Data Preprocessing}%
\label{sec:preprocessing}%
After text extraction, we preprocessed each document to isolate the
main body content. We removed references and appendices by discarding
all text after the final occurrence of the keyword ``References'' in
each document. We then applied regular expressions to strip inline
citations (e.g., bracketed reference numbers), meta-information such as
page headers and footers, and extraneous line breaks. Finally, we
converted all text to lowercase to ensure uniform input for the
language models. The resulting preprocessed corpus contains
879~documents.

\subsection{Two-step Approach for Cost-Effective Mining of Insights (RQ1)}%
\label{sec:Information-extraction}%

We used a two-step approach for cost-effective insight mining. In the first step, we extracted candidate challenges from each paper with GPT-3.5 (gpt-3.5-turbo-0301). In the second step, we used GPT-4 (gpt-4-0314) to deduplicate and retain the five most important challenges per paper.

The next two sections describe our prompt design for each step. We conducted all prompt experiments in Jupyter notebooks with the OpenAI Python Library (v0.27.4) rather than in OpenAI’s web interface. We set the temperature parameter to zero for all experiments, which produces deterministic outputs and is recommended for factual tasks such as data extraction \citep{prompting-best-practices}. This setting also made prompt iterations directly comparable, because identical inputs always produced identical outputs. We left the top-p parameter at its default value, as recommended by OpenAI.

\subsubsection{Step 1: Insight mining with GPT-3.5}%
\label{sec:method:step1}%
We experimented with various prompts to extract research challenges from the text corpus. Our prompt design drew on best practices from both scholarly and gray literature \citep{2312.16171v1.pdf,2210.09150.pdf,2304.07619.pdf,openai_promptdesign,2305.14930.pdf,taxonomy}. For each modification, we evaluated the output on a sample of documents and retained changes only when they improved the results. This iterative process produced the following prompt:
\begin{quote}%
\textit{%
 Forget all the above.
 You are InfoMinerGPT, a language model trained to extract challenges for researchers in the field of HCI from academic papers.
 You output a list of challenges, one per line, without explanations.
 If there are no challenges mentioned in the text, you output 'None'.
 Here is the academic paper:{\textbackslash}n{\textbackslash}n
}%
\end{quote}%
The first part of this prompt instructs GPT-3.5 to disregard its default instructions and persona \citep{2211.09527.pdf}. Because GPT-3.5 is designed as a general-purpose AI assistant that provides explanations, overriding its default behavior with an expert persona improves task-specific performance \citep{2305.14930.pdf,2302.11382.pdf,2211.09527.pdf,2312.16171v1.pdf} and reduces verbosity. Similar prompt instructions have been used in other studies (e.g., \citet{2304.07619.pdf}). We also tested alternative formats, such as instructing the model to ``never break character'' and few-shot prompting (i.e., providing examples in the prompt) \citep{2005.14165.pdf}, but the above prompt consistently produced plausible research challenges relevant to the source text.
%
\subsubsection{Step 2: Information filtering with GPT-4}%
\label{sec:method:step2}%
In the second step, we used GPT-4 to filter and rank the candidate challenges from Step~1. We appended each paper's list of extracted challenges to the following prompt:
\begin{quote}%
\textit{%
 Below is a JSON list of challenges extracted from an academic paper.
 Please remove any duplicates.
 Then keep only the 5 most important challenges for researchers in the field of HCI.
 Select research challenges based on their relevance to current HCI trends, potential for future impact, and novelty. 
 You output a list of challenges, one per line, without explanations.
 Here are the challenges:{\textbackslash}n{\textbackslash}n
}%
\end{quote}%
This prompt required less design iteration than the Step~1 prompt, as GPT-4 followed instructions more reliably \citep{gpt-4.pdf,2304.13712.pdf}. Computational priming with a persona \citep{PRIMING} was not necessary for this step. The three selection criteria (relevance to current HCI trends, potential for future impact, and novelty) were chosen to prioritize challenges that reflect the field's active research directions while also surfacing emerging areas not yet widely addressed. These criteria align with common frameworks for assessing research significance in HCI, where contribution, significance, and novelty are standard dimensions of evaluation (e.g., the ACM CHI review criteria).
%
\subsubsection{Summary of prompt design}%
\label{sec:prompt-design-summary}%
The prompts produced consistent results across our experiments. However, LLM outputs are sensitive to prompt design (see Section~\ref{sec:limitations}), and we do not claim that our prompts represent an optimal solution for uncovering ground truth. Rather, our goal was to extract signal from noise, and the iterative prompt design process reduced noise with each iteration. We evaluate the quality of the extracted information in Section~\ref{sec:evaluation}.
%
\subsubsection{Quality Assurance}%
\label{sec:quality-assurance}%
We took several measures to ensure the quality and consistency of the extracted data. First, we developed prompts iteratively in Jupyter notebooks, tested each modification on a sample of documents, and retained changes only when they improved output quality (Section~\ref{sec:method:step1}). Second, we set the temperature parameter to zero for all API calls to produce deterministic outputs, which made systematic comparison across prompt iterations possible. Third, the two-step pipeline provides built-in redundancy: GPT-3.5 performs broad extraction in Step~1, and GPT-4 filters and ranks the results in Step~2, which reduces the likelihood that low-quality or irrelevant challenges persist in the final output. Fourth, we used string matching to identify artifacts in the Step~1 output and found that only 0.9\% of extracted statements were noise, which GPT-4 automatically discarded during Step~2 (Section~\ref{sec:noise}). Finally, we conducted a human evaluation in which two researchers independently reviewed 45~sampled papers to assess the alignment between machine-extracted and human-identified challenges (Section~\ref{sec:evaluation}). Together, these measures address quality at multiple stages of the pipeline, from prompt design through final evaluation.
%
\subsubsection{API querying statistics and costs}%
\label{sec:api-stats}%
We split the source text into sentences with NLTK's sentence tokenizer and queried the API in batches. The API could handle most papers in batches of 90~sentences without exceeding the model's maximum context length of 4,096~tokens, but papers with very long sentences caused an ``InvalidRequestError.'' To address this, we implemented a fallback that reduced the batch size by 20~sentences when such errors occurred. Most papers were processed with a batch size of 90~sentences, while a few required a batch size as low as 50~sentences.

The extraction required cloud-based computational resources, with a total processing time of 17.5~hours (13~hours and 45~minutes for Step~1 and 3~hours and 51~minutes for Step~2). We monitored the process and occasionally restarted it when a ``RateLimit’’ error indicated that OpenAI’s API was overloaded. The total cost for all experiments, prompt design, and batch processing was US~\$49.95. The extraction scripts and Jupyter notebooks are available in the OSF repository for replication.%
%
\subsection{Topic Modeling}%
\label{sec:Topic-Modeling}%
We analyzed the extracted research challenges with a topic modeling approach. As a first step, we converted the challenge statements into sentence embeddings, which are high-dimensional vector representations that capture the semantic and syntactic properties of linguistic units. We compared three embedding models for this task: the all-mpnet-base-v2 pre-trained Sentence-BERT model \citep{1810.04805.pdf}, OpenAI’s text-embedding-ada-002, and the CLIP model \citep{CLIP}. Sentence-BERT produced the most coherent clusters, with a silhouette score of 0.27, compared to 0.19 and 0.24 for text-embedding-ada-002 and CLIP, respectively.

We then clustered the embeddings with BERTopic \citep{grootendorst2022bertopic}, a semi-automated topic modeling approach that combines BERT \citep{1810.04805.pdf} for embedding generation, UMAP \citep{UMAP} for dimensionality reduction, HDBSCAN \citep{HDBSCAN} for clustering, and c-TF-IDF \citep{cTFIDF} for topic creation. BERTopic represents each topic as a bag of words, which we initially reviewed manually to assign descriptive labels. For many topics, this labeling was straightforward, and inter-rater agreement was not required \citep{McDonald_Reliability_CSCW19.pdf}. The manual review helped us understand the data and assess topic quality. When we later optimized UMAP, HDBSCAN, and BERTopic parameters, we switched to automatic labeling with GPT-3.5 so that we could iterate over parameter configurations more efficiently. After this iterative exploration, the first author manually labeled the final set of 113~topics.

We chose BERTopic over session-based groupings for two reasons. First, some CHI sessions have generic names (e.g., ``Building Bridges'' and ``Discovery Track Monday'') that do not serve as effective thematic groupings for research challenges. Second, the challenges mentioned in a paper are not always related to its main topic or session, so session titles may not accurately represent the research challenges discussed.

\subsection{Interactive Visualization}%
\label{sec:visualization}%
We projected the high-dimensional embeddings into two-dimensional space to create an interactive visualization of the clustered research challenges. We compared two dimensionality reduction algorithms for this projection: t-distributed Stochastic Neighbor Embedding (t-SNE) \citep{TSNE} and Uniform Manifold Approximation and Projection (UMAP) \citep{UMAP}. We chose UMAP because it better preserves global structure \citep{Understanding-UMAP,UMAP}.

\begin{figure}[!htb]%
\centering%
\includegraphics[width=.9\linewidth]{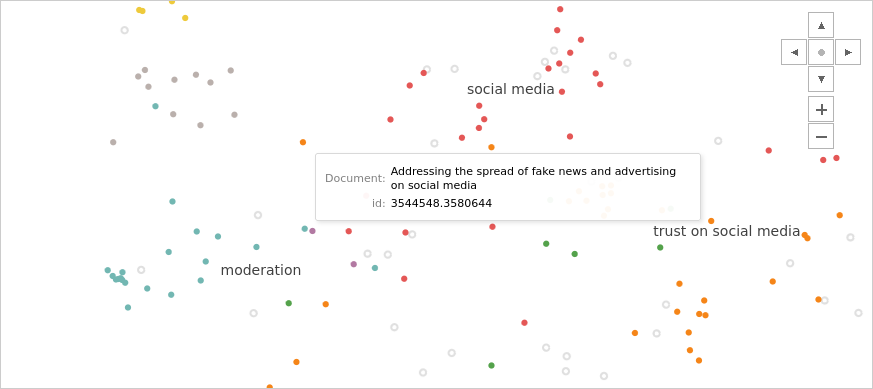}
\caption{Zoomed-in view of the interactive visualization with its light-weight user interface. Each dot represents a research challenge extracted from the CHI 2023 proceedings, color-coded by topic cluster. Hovering over a dot reveals the challenge statement and the DOI of the source paper. Cluster labels identify thematic groupings such as ``social media'' and ``misinformation.'' The full interactive visualization is available at \href{https://HCI-research-challenges.github.io}{https://HCI-research-challenges.github.io}.}
\label{fig:screenshot}%
\end{figure}

The resulting interactive visualization is available at \href{https://HCI-research-challenges.github.io}{https://HCI-research-challenges.github.io}. We created it with Vega-Altair (version 4.2.2), a Python library for declarative, interactive data visualizations based on the Vega and Vega-Lite grammars \citep{Altair}. Vega-Altair generates a Vega-Lite specification in which each data point encodes the two-dimensional UMAP coordinates, the cluster label, the challenge statement, and the DOI of the source paper. The specification and the Vega rendering libraries are exported as a self-contained HTML file and hosted as a static page on GitHub Pages, with no server-side processing required. \autoref{fig:screenshot} shows a zoomed-in view of the visualization.

Each research challenge appears as a dot, color-coded by topic cluster. UMAP positions thematically related clusters near each other (e.g., the social media clusters in \autoref{fig:screenshot}). Challenges that UMAP did not assign to any cluster appear as light gray dots. Cluster labels are centered within each cluster.

Users can zoom with the scroll wheel, pan by dragging the canvas, and hover over a dot to see the research challenge statement and the DOI of the source paper. To illustrate, a researcher interested in privacy could zoom into the ``privacy and security'' cluster ($n = 174$), hover over individual challenges to read their statements, and follow DOI links to consult the original papers. Similarly, an HCI educator designing a graduate seminar could use the overview to identify which research themes are most active, compare the relative sizes of topic clusters, and select underexplored areas as discussion topics for students. All prompts used in the extraction pipeline are documented in Appendix~B. The full dataset, extraction scripts, and example outputs are available in the OSF repository.\footnote{\href{https://osf.io/fbsq7/?view_only=b6b86e89fe444a74a33950b452a55242}{https://osf.io/fbsq7/?view\_only=b6b86e89fe444a74a33950b452a55242}}

\section{Research Challenges in HCI}%
\label{sec:results}%
\autoref{fig:viz} shows the research challenges identified through our pipeline. Readers can explore the full results through the interactive visualization (\href{https://hci-research-challenges.github.io}{https://hci-research-challenges.github.io}) and the dataset (\href{https://osf.io/fbsq7/?view\_only=b6b86e89fe444a74a33950b452a55242}{https://osf.io/fbsq7/?view\_only=b6b86e89fe444a74a33950b452a55242}). We conclude by examining how these challenges relate to HCI's grand challenges (Section~\ref{sec:grand}).

\input{FIG-CHALLENGES}%

\subsection{Current Research Challenges and Themes (RQ2)}
GPT-3.5 extracted 34,638~challenges from 879~papers, with an average of 39.3~challenges per paper ($SD = 21.5$; $Min = 3$; $Max = 198$) and a mean length of 10.7~tokens per challenge ($SD = 4.9$).
GPT-4 then filtered this set to 4,392~challenges by selecting 3 to 5 per paper ($M = 4.99$, $SD = 0.06$), with an average length of 11.1~tokens ($SD = 4.3$).
\autoref{tab:sample} and \autoref{tab:example} show examples of the resulting challenges.

\begin{table}[htb]%
\caption{Random sample of research challenges extracted from the CHI 2023 proceedings.}%
\label{tab:sample}%
\small%
\begin{tabularx}{\textwidth}{
    >{\hsize=.75\hsize}X
    >{\hsize=.25\hsize}X
}%
\toprule%
    Challenge & Paper \\
\midrule%
    Investigating how embodying physics-aware avatars impacts task performance in VR. & \citet{3544548.3580979} \\
    Developing automated tools to detect and categorize OOD data in NLP models & \citet{3544548.3580741} \\
    Understanding how designers make sense of the workings of physical interfaces built with CV markers & \citet{3544548.3580643} \\
    Improving findability and actionability of privacy controls for online behavioral advertising & \citet{3544548.3580773} \\
    Dealing with nuanced ethical concerns in conducting HCI research in the wild & \citet{3544548.3580875} \\
\bottomrule%
\end{tabularx}%
\end{table}%

BERTopic grouped the research challenges into 113~topics (see Appendix A, \autoref{tab:challenges}).
The largest topic clusters address privacy and security ($n = 174$), virtual reality ($n = 139$), interaction ($n = 107$), marginalized communities ($n = 106$), and improving accuracy (e.g., object detection; $n = 103$).
The topics span both traditional HCI concerns (interaction, gestures, accessibility, colors, cognitive biases, UX~design, and user interfaces) and emerging areas such as human-AI collaboration, LLM-assisted writing, AI agents, chatbots, and virtual/augmented reality.

\begin{figure}[htb]%
  \centering%
  \begin{minipage}{0.475\linewidth}
  \includegraphics[width=\linewidth]{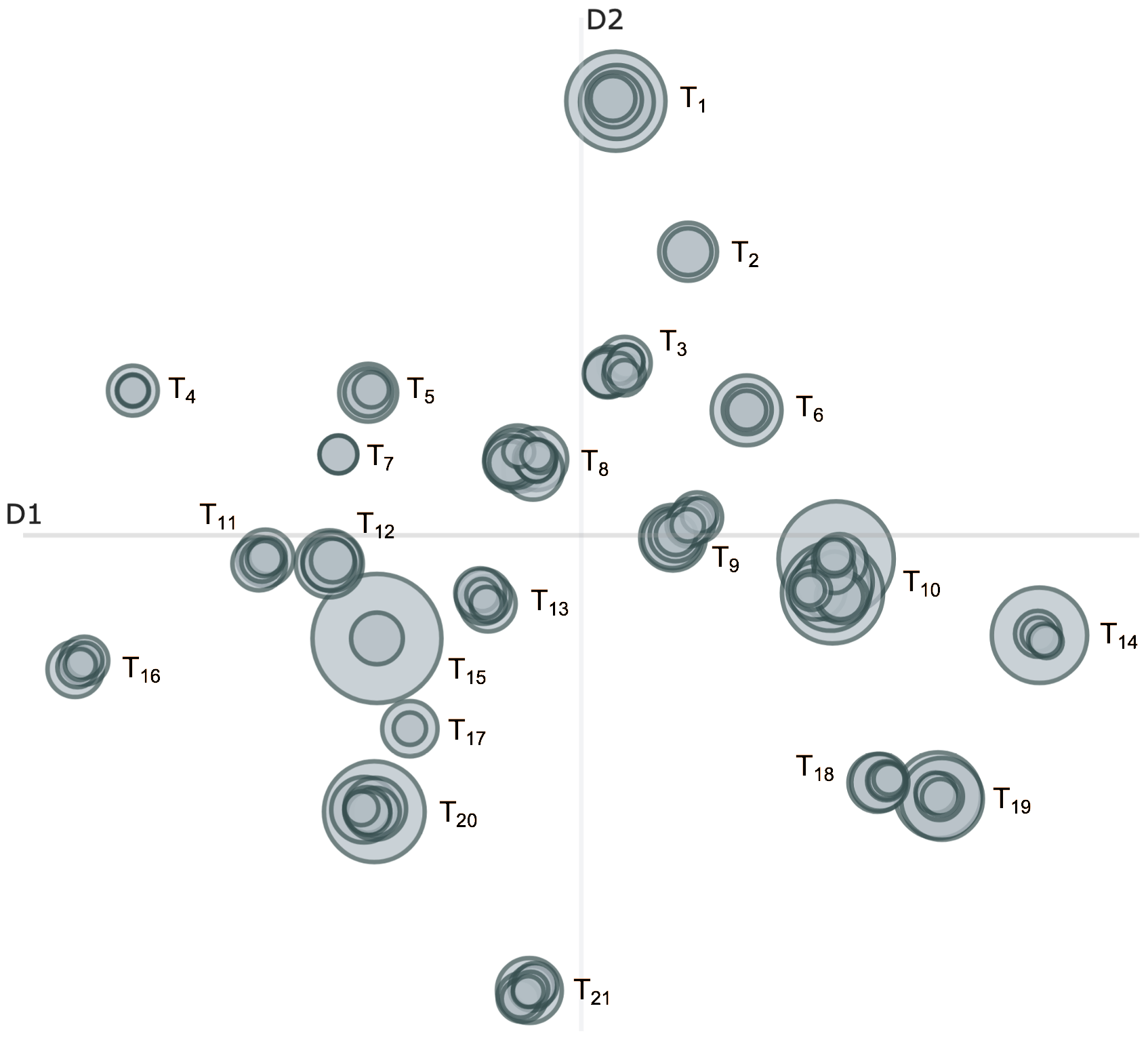}%
  \end{minipage}%
  ~
  \begin{minipage}{0.475\linewidth}
    \begin{tcolorbox}[rounded corners,colback=white,boxrule=.5pt, left=3mm, right=1mm, top=1mm, bottom=1mm]
    \scriptsize%
    \raggedright%
    \begin{itemize}[leftmargin=11pt, itemsep=0pt, parsep=0pt, topsep=0pt, partopsep=0pt]%
    \item[ T\textsubscript{1} ] Gestures, eye/gaze tracking
    \item[ T\textsubscript{2} ] Voice assistants, speech recognition
    \item[ T\textsubscript{3} ] Visualization, storytelling
    \item[ T\textsubscript{4} ] Conversational AI, chatbots
    \item[ T\textsubscript{5} ] Writing, nlp, language models
    \item[ T\textsubscript{6} ] Notifications, fitness tracking, emotions
    \item[ T\textsubscript{7} ] Recommendations, decision-making
    \item[ T\textsubscript{8} ] Human-AI collaboration, XAI and transparency, trust
    \item[ T\textsubscript{9} ] HCI research, learning analytics, remote work, conflicts
    \item[ T\textsubscript{10} ] VR/AR/mixed reality, remote work
    \item[ T\textsubscript{11} ] Games, interaction, engagement, rituals
    \item[ T\textsubscript{12} ] Misinformation, social media
    \item[ T\textsubscript{13} ] Data science, scalable datasets, health data
    \item[ T\textsubscript{14} ] Haptic/tactile feedback, sensing, stimulation
    \item[ T\textsubscript{15} ] Privacy and security, consent
    \item[ T\textsubscript{16} ] Ethics, algorithmic fairness, biases
    \item[ T\textsubscript{17} ] Digital literacy, family and parenting
    \item[ T\textsubscript{18} ] Wearables, textiles, fabrication
    \item[ T\textsubscript{19} ] Blind/visually impaired, older adults, DHH
    \item[ T\textsubscript{20} ] Marginalized communities, LGBTQ, social norms
    \item[ T\textsubscript{21} ] Health, well-being
    \end{itemize}
    \end{tcolorbox}
  \end{minipage}
  \caption{Intertopic distance map with 21 research themes (\mbox{T\textsubscript{1}}--\mbox{T\textsubscript{21}}).
  Each circle on the map represents a topic identified by BERTopic in the CHI 2023 proceedings.
  The size of the topics reflects the amount of research challenges associated with that topic.
  Topics that are close to each other can be considered to have similar content or themes, while topics far apart are more distinct or less related. 
  Topics that are similar to each other are placed closer together, while dissimilar topics are farther apart.
  A topic cluster forms a research theme within HCI.
  We represent the research themes with selected salient keywords occurring in the topics that make up the theme.}%
  \label{fig:intertopic-map}%
\end{figure}

BERTopic also produces an intertopic distance map (\autoref{fig:intertopic-map}), which represents the relationships between topics in a two-dimensional UMAP space. From this map, we identify at least 21~main research themes (T\textsubscript{1}--T\textsubscript{21}) in HCI. Because BERTopic’s hyperparameters affect the number and boundaries of topics, this count should be treated as approximate rather than definitive.

The intertopic distance map shows no strong outliers among HCI’s research themes. Most themes are well-connected, and several overlap. For example, `privacy and security’ (T\textsubscript{15}) clusters near `misinformation, social media’ (T\textsubscript{12}), `data science, health data’ (T\textsubscript{13}), and `Human-AI collaboration, XAI, transparency and trust’ (T\textsubscript{8}). `VR/AR/mixed reality’ (T\textsubscript{10}) and `haptic/tactile feedback’ (T\textsubscript{14}) also occupy adjacent positions. Other theme pairs are more distant: `gestures, eye/gaze tracking’ (T\textsubscript{1}) and `health, well-being’ (T\textsubscript{21}), `haptic feedback, sensing, stimulation’ (T\textsubscript{14}) and `ethics’ (T\textsubscript{16}), and `conversational AI’ (T\textsubscript{4}) and `blind/visually impaired’ (T\textsubscript{19}). These distances may point to underexplored connections between subfields.


We now compare the identified research challenges and themes with HCI's grand challenges.

\subsection{Comparison with HCI's Grand Challenges and UN SDGs (RQ3)}

\input{FIG-GC-COMPARISON}
A qualitative comparison of the grand challenges from Section~\ref{sec:grand} with the extracted research challenges suggests a difference in granularity. The grand challenges define broad strategic domains, whereas the extracted challenges address specific subproblems within those domains.

\subsubsection{Method}%
We embed the grand challenges from prior literature into the same UMAP space \citep{UMAP} as the extracted research challenges, which allows direct visual comparison (\autoref{fig:gccomparison}). Because embedding positions are sensitive to surface-level lexical overlap, we made two adjustments. First, we separated compound grand challenges from \citet{10.1080@10447318.2019.1619259.pdf} into distinct items (e.g., ``Creativity and Learning'' became ``Creativity'' and ``Learning''). Second, we removed the word ``sustainable'' from several SDG labels, as its repeated occurrence caused the SDGs to cluster together regardless of their thematic content. The modified labels are listed in \autoref{fig:gccomparison}.

\subsubsection{Results and discussion}



\autoref{fig:gccomparison} depicts the extracted research challenges alongside the grand challenges and the UN's Sustainable Development Goals (SDGs). Each grand challenge has at least one nearby cluster in the embedding space, which suggests that these established priorities remain well-represented in current HCI research. The grand challenge of ``Ethics, Privacy and Security'' \citep{10.1080@10447318.2019.1619259.pdf} corresponds to the largest cluster in our data (privacy and security, $n = 174$) and to related clusters on transparency in algorithms ($n = 15$) and ethics and ethical guidelines ($n = 32$). ``Well-being, Health and Eudaimonia'' maps to five clusters (clinical AI, $n = 37$; mental wellbeing, $n = 25$; chatbots and mental health, $n = 26$; health data, $n = 32$; digital health interventions, $n = 10$) for a total of 130~challenges. ``Accessibility and Universal Access'' aligns with clusters on accessibility for disabilities ($n = 77$), blind and visually impaired users ($n = 70$), and DHH accessibility ($n = 14$) for a total of 161~challenges. The grand challenge of ``Creativity'' finds a counterpart in creativity support tools ($n = 20$), and \citeauthor{p24-shneiderman.pdf}'s ``Promote lifelong learning'' maps to learning and group awareness ($n = 26$) and learning analytics ($n = 13$).

Several prominent clusters fall outside the scope of both grand challenge frameworks. Human-AI collaboration ($n = 46$), writing assistants ($n = 35$), and AI agents ($n = 28$) represent concerns about human-AI interaction that these frameworks did not anticipate. AI design ideation ($n = 10$) and explainable AI ($n = 17$) point to a related shift: AI has become both a design material and a subject of inquiry in HCI. Taken together, these clusters suggest that the field's research priorities have evolved beyond the existing grand challenge frameworks, and that these frameworks may benefit from revision to account for the ways humans collaborate with, rely on, and respond to increasingly capable AI systems (see also Section~\ref{sec:hci-implications}).

Additional alignments emerge across the remaining grand challenges. \citeauthor{p24-shneiderman.pdf}'s call to ``design novel input and output devices'' maps to haptic feedback ($n = 95$), gestures ($n = 57$), and textile interfaces ($n = 16$), which together constitute one of the most active research threads in the corpus. ``Support successful aging strategies'' corresponds to older adults and care ($n = 23$), and ``shift from user experience to community experience'' resonates with clusters on marginalized communities ($n = 106$), LGBTQ and sexual consent ($n = 40$), and social norms ($n = 20$), although the emphasis has shifted from community-level UX toward equity and social justice. \citeauthor{10.1080@10447318.2019.1619259.pdf}'s ``Social Organization and Democracy'' aligns with trust and moderation in social media ($n = 47$) and misinformation in news ($n = 29$), clusters that reflect the political dimensions of platform design in the era of large-scale social platforms.

Other grand challenges have little direct representation in our data. \citeauthor{p24-shneiderman.pdf}'s ``engineer new business models'' and ``refine theories of persuasion'' lack obvious cluster matches, though the latter may have evolved in an unexpected direction: deceptive design ($n = 28$) addresses the manipulative application of persuasive techniques. This shift suggests that HCI research has reframed persuasion as a problem to be countered rather than a capacity to be developed. ``Encourage resource conservation'' maps only indirectly to environment and eco feedback ($n = 45$). These gaps suggest that current HCI research gravitates toward challenges where technology-user interaction is most immediate, while more systemic or organizational challenges remain underexplored in the CHI 2023 proceedings.

The UN's SDGs remain closely grouped in the embedding space even after our lexical modifications, which reflects their shared thematic orientation. A few extracted clusters show direct thematic alignment: waste management ($n = 12$) and food and food delivery ($n = 24$) correspond to SDGs on responsible consumption and zero hunger, while the environment and eco feedback cluster ($n = 45$) relates to climate action and sustainable communities. Beyond these, most areas of current HCI research connect to sustainability only indirectly, and some show no apparent connection.

This partial misalignment suggests an opportunity. HCI's interdisciplinary nature positions it well to contribute to goals related to health equity, education, and sustainable infrastructure, yet the CHI 2023 research agenda does not systematically engage with these objectives. We do not argue that all HCI research should orient toward sustainability, but greater awareness of the SDGs may open productive research directions and strengthen the societal relevance of HCI work.

\section{Evaluation}%
\label{sec:evaluation}%
We evaluate the extracted challenges from two angles: quantitative metrics that situate our results relative to standard NLP baselines (Section~\ref{sec:metrics}), and qualitative assessments that examine whether the LLMs produced valid, relevant research challenges (sections~\ref{sec:eval:task} and~\ref{sec:eval:quality}).

\subsection{Quantitative Evaluation}%
\label{sec:metrics}%
No established benchmark exists for extracting research challenges from scholarly text, so our primary evaluation is qualitative. We nonetheless report several quantitative metrics to characterize the relationship between the LLM outputs and their source texts.

Three metrics capture surface-level similarity between extracted challenges and the source text. Exact Match (EM) \citep{bulian2022tomayto} measures the percentage of instances where two strings are identical. A high EM score would suggest that the LLMs copied statements directly from the source, while a low score indicates rephrasing. Edit Distance (ED) counts the minimum insertions, deletions, or substitutions needed to transform one string into another, and Word Error Rate (WER) quantifies word-level errors in a generated sequence normalized by the reference length. Together, these metrics indicate the extent to which the LLMs paraphrased the source material rather than reproducing it verbatim.

We also report several standard NLP metrics: BLEU \citep{1073083.1073135.pdf} for n-gram precision between generated and reference texts; ROUGE-1 and ROUGE-L \citep{was2004.pdf} for unigram overlap and longest common subsequences; METEOR \citep{METEOR} for synonym-aware and order-sensitive comparison; BLANC \citep{BLANC} for summary quality; and BERT\-Score \citep{1904.09675.pdf} for embedding-based cosine similarity. We include these metrics primarily as reference points for future work.

Because no gold-standard reference set exists for our task, we approximated one by pairing each extracted challenge with its closest match in the source text based on embedding cosine similarity. In Step~1, we paired each GPT-3.5 challenge with the most similar sentence in the source document. In Step~2, we paired each GPT-4 statement with the most similar statement from the GPT-3.5 output. \autoref{tab:metrics} reports the results.

\input{TAB-METRICS}%

As discussed in Section~\ref{sec:evaluatingllms}, these metrics were designed for other tasks: BLEU, WER, and METEOR evaluate machine translation, and BLANC scores text summaries. They measure surface-level textual overlap but do not assess whether an extracted statement constitutes a valid research challenge. We therefore complement these quantitative baselines with the qualitative evaluation in sections~\ref{sec:eval:task} and~\ref{sec:eval:quality}.

\subsection{Task-specific Human Evaluation}%
\label{sec:eval:task}%
The central question for this evaluation is whether the LLMs produced genuine research challenges rather than arbitrary sentences from the source documents. Specifically, we ask whether GPT-3.5 extracted statements that qualify as research challenges and whether GPT-4 then selected the most relevant ones from each GPT-3.5 list. Because no ground truth exists for this task, we rely on two forms of grounding: `narrow grounding' compares the LLM output against statements in the supplied prompt, while `broad grounding' draws on our own domain knowledge to judge the outputs. We apply both forms to a random sample of documents, as detailed below.
%
\subsubsection{Qualitative analysis methodology}%
\label{sec:qualeval}%
A Postdoctoral Researcher with six years of HCI experience independently reviewed a random sample of 45~papers (approximately 5\% of the corpus) and identified the five most important research challenges from each paper without access to the LLM-generated lists. Two authors then compared these human-generated lists with the corresponding GPT-4 lists. We initially treated lists as equivalent when three or more GPT-4 statements matched the human-generated list lexically or semantically, though we later relaxed this criterion (see Section~\ref{sec:qualeval:prelim}). We accepted partial matches when the overall meaning was preserved. We also assessed whether the GPT-4 lists contained plausible HCI research challenges, adopting a conservative threshold: a statement counted as a research challenge if it could reasonably be pursued as one, even outside our own areas of expertise. We measured inter-rater agreement with Cohen’s kappa \citep{cohen}.
%
%
%
\subsubsection{Preliminary observations and adjustment of methodology}%
\label{sec:qualeval:prelim}%
The human annotator performed the same two-step process as the LLMs: first compiling a list of challenges from each paper (analogous to the GPT-3.5 task) and then selecting the five most important ones (analogous to the GPT-4 task). We deliberately included papers outside the PostDoc’s area of expertise to test the LLMs across HCI subdomains and to mirror situations common in academic practice, such as peer review of work outside one’s specialty.

The task proved demanding for the human annotator. In unfamiliar areas such as embodied interaction, the annotator tended to copy verbatim statements from the paper rather than formulate them as research challenges. The cognitive load was especially high when the annotator had to narrow each list to five statements.

\input{TAB-EXAMPLE}%

Because the narrowing step yielded diminishing returns and because simple concatenation of list items could have artificially reduced the list, we adopted a permissive evaluation strategy. Rather than requiring the human list to contain exactly five challenges, we compared the full set of human-provided challenges (between 4 and 21 per paper) against the GPT-4 list.

\subsubsection{Alignment with human judgment}%
\label{sec:qualeval:results}%

Both raters agreed with near-perfect reliability ($\kappa = 0.97$) that the extracted statements represent plausible HCI research challenges. Validating each individual challenge would require expertise across all HCI sub-fields, but the high agreement indicates that the statements are, at minimum, face-valid.

Inter-rater agreement on the alignment between GPT-4 and human-identified challenges was also high ($\kappa = 0.86$). In about two-thirds of the sampled papers (65.9\% and 63.6\% for the two raters), we found a full match between the human-identified and GPT-4-extracted challenges. Three factors account for the remaining discrepancies. First, the PostDoc focused on each paper's abstract, introduction, and discussion, whereas the language model processed the full text without such selectivity. Second, the human annotator experienced fatigue over the course of the task and at times resorted to scanning headlines and copying verbatim statements, which may have caused the annotator to overlook some challenges. Third, the human-created lists typically centered on each paper's narrow main topic, while GPT-4 sometimes identified broader challenges. In \citet{3544548.3581503.pdf}, for instance, the human-extracted challenges focused on ``designing and deploying LLM-driven chatbots for public health intervention,'' whereas GPT-4 also included broader challenges such as ``controlling the output of LLM-based chatbots to avoid unintended or biased responses'' and ``ensuring privacy and security of user data in AI-powered interventions.'' In other cases, GPT-4 identified challenges that the human annotator missed entirely. In \citet{3544548.3581393}, for example, GPT-4 included ``mitigating risks of de-skilling clinicians in the use of AI,'' which did not appear in the human list.

The one-third of cases that did not fully align does not imply that the GPT-4 lists contain invalid challenges. Rather, the human and the model emphasized different aspects of each paper. In several of these cases, the GPT-4 lists were more complete than the human-created lists in both lexical and semantic terms.

We summarize the key performance indicators for the extraction pipeline. At the task validity level, both raters agreed with near-perfect reliability ($\kappa = 0.97$) that the extracted statements represent plausible HCI research challenges. At the alignment level, inter-rater agreement was high ($\kappa = 0.86$), with a mean full-alignment rate of 64.8\% across the two raters (65.9\% and 63.6\%, respectively) on the 45-paper sample. The remaining cases reflected differences in emphasis rather than extraction errors. At the pipeline quality level, noise in Step~1 was minimal: only 0.9\% of GPT-3.5 outputs were non-challenge artifacts. Semantic fidelity was high, as indicated by a mean cosine distance of $d = 0.075$ ($SD = 0.066$) between extracted challenges and their best-matching source sentences. In Step~2, GPT-4 reproduced 98.6\% of selected challenges verbatim from the GPT-3.5 output, which confirms that the filtering step introduced negligible paraphrasing. Together with the NLP metrics in \autoref{tab:metrics}, these indicators suggest that the two-step pipeline produces outputs that are both valid as research challenges and faithful to the source text.

\subsubsection{Other outstanding observations on LLM performance}%
\label{sec:qualeval:observations}%
%
Our review of the 45 sampled papers yielded three additional observations about LLM performance.

First, acronyms were prevalent in the challenges extracted by both GPT-3.5 and GPT-4, which reflects their widespread use in the source papers. While some acronyms are broadly familiar to HCI researchers (e.g., AI, VR, UX, NLP), others are specific to sub-communities and may not be immediately interpretable: ES (Environmental Sustainability), SHCI (Sustainable HCI), RAI (Responsible AI), DHH (d/Deaf and Hard-of-Hearing), AAC (Augmentative and Alternative Communication), and OCCs (Online Critique Communities).

Second, some challenges lacked sufficient context for accurate interpretation and at first appeared to be hallucinations. For example, the GPT-4 statement ``Addressing scalability issues for large-scale text data'' seemed fabricated but actually referred to a scalability issue discussed in the limitations section of \citet{3544548.3580741}. The statement is a valid research challenge, yet it is difficult to interpret without the original paper's context.

Third, GPT-4's selection of research challenges sometimes alternated between topic-specific and broad statements, a pattern that differed from the human-extracted lists. In \citet{3544548.3580907.pdf}, for instance, GPT-4 identified topic-specific challenges such as ``Supporting journalists in exploring multiple angles for a given press release'' and ``Ensuring responsible journalism with LLM-generated angles,'' alongside broader ones such as ``Addressing limitations of publicly available LLMs, such as outdated world knowledge and bias in training data'' and ``Mitigating bias in LLM-generated implications.'' The latter was flagged as a potential hallucination. However, \citet{3544548.3580907.pdf} discuss bias mitigation during prompt design when generating ``implications'' from text, even though the exact phrase does not appear in the paper. Across all 45 sampled papers, we found no confirmed evidence of hallucinations in the extracted challenge sets.
%
%
\subsection{Evaluation of Performance}%
\label{sec:eval:quality}%
Beyond the alignment with human judgment established above, we assess two further aspects of output quality: the presence of noise in Step~1 (Section \ref{sec:noise}) and the semantic similarity between the extracted challenges and the source text (Section \ref{sec:similarity}).

\subsubsection{Noise in GPT-3.5 results}%
\label{sec:noise}%
We manually reviewed the challenges extracted by GPT-3.5 and identified non-research statements such as:
\begin{itemize}
    \item 
``Challenges for researchers in the field of HCI from this paper:``
    \item 
``Challenges not mentioned in the text.``
\end{itemize}
These statements are not research challenges but artifacts of GPT-3.5's tendency to produce meta-commentary alongside its output. Basic string matching and regular expressions can filter them out. Only 317 of 34,638 statements (${\approx}0.9\%$) fell into this category in Step~1. In Step~2, GPT-4 ignored these artifacts when selecting research challenges from the GPT-3.5-provided list.

CHI papers frequently contain ligatures (e.g., ``ﬀ'') that introduce multibyte characters into the extracted source text, sometimes producing white space within words. Neither GPT-3.5 nor GPT-4 had difficulty processing these typographic artifacts.

\subsubsection{Semantic similarity}%
\label{sec:similarity}%
An information extraction pipeline should produce output that is semantically close to the source text. To assess this property, we measured embedding cosine distances between extracted statements and source sentences. We split each source document into sentences with the NLTK sentence tokenizer and embedded them with the all-mpnet-base-v2 pre-trained sentence transformer model. For each GPT-3.5 statement, we then identified the best-matching source sentence by cosine similarity. In Step~1, we expected a long-tailed distribution of cosine distances, in which most challenges closely match a source sentence and few show substantial divergence. We applied the same procedure in Step~2 to compare GPT-4 statements against the GPT-3.5 output.

\begin{figure}[htb]%
\centering%
         \includegraphics[width=.5\textwidth]{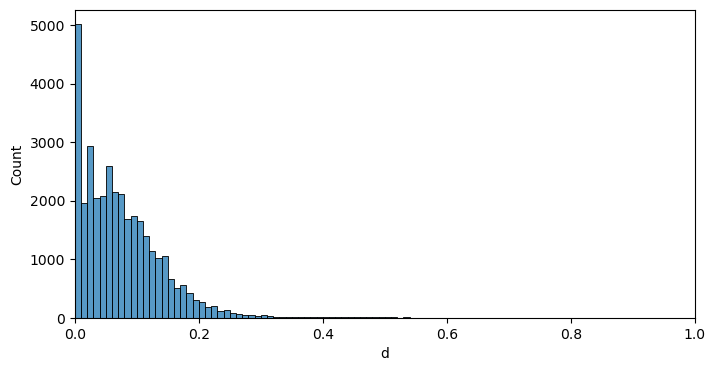}
\caption{Cosine distances $d$ between the embedded statements extracted by GPT-3.5 in Step 1 and the best-matching sentence embeddings from the source text.
Most statements are very similar to a sentence in the source text, and only a few show substantial divergence. A high cosine distance does not necessarily indicate an invalid research challenge, because GPT-3.5 may have paraphrased or summarized the source text.%
}%
\label{fig:distances}%
\end{figure}%

\autoref{fig:distances} shows the embedding cosine distance ($d$) for the best-matching pairs, where $d = 0$ indicates identity with a source sentence and lower values indicate greater semantic similarity. In Step~1, the GPT-3.5-extracted challenges were generally close to the source text, with a mean cosine distance of $d = 0.075$ ($SD = 0.0662$, $Q1 = 0.0252$, $Q2 = 0.0628$, $Q3 = 0.1093$).

In Step~2, we expected GPT-4 to reproduce its selections verbatim, since the prompt instructed the model to choose the five most important challenges from the GPT-3.5 list rather than to rephrase them. The results confirmed this expectation: GPT-4 copied 4,333 statements verbatim from the Step~1 output and altered the wording in only 60 cases (1.38\%).

\section{Discussion}%
\label{sec:discussion}%
The application of large language models to academic research opens new possibilities for large-scale data analysis and insight generation, with potential consequences that extend well beyond HCI. In what follows, we consider how LLMs fit into existing research practice (Section~\ref{sec:LLMresearch}), what cost efficiencies they offer (Section~\ref{sec:cost-efficiency}), where their limitations lie (Section~\ref{sec:reviewerbelief}), and what implications arise for HCI practice, education, and policy (Section~\ref{sec:hci-implications}).

\subsection{Applying LLMs for Insight Mining at Scale}%
\label{sec:LLMresearch}%
%
Foundation models \citep{Bommasani2021FoundationModels} are pre-trained, general-purpose systems that handle a wide range of tasks through prompting, although their effectiveness varies by task. For research questions previously considered intractable at scale, in-context learning and careful prompt design open new avenues for exploration. Large language models add a significant capability to the researcher’s toolkit, and the shift is already visible outside academia, where numerous applications have emerged under the developing framework of \textit{LMOps} or \textit{LLMOps} \citep{LLMOps}. LLM-based agentic systems such as Auto-GPT \citep{AutoGPT}, for example, use GPT-4 for autonomous problem-solving and have attracted widespread attention, as reflected in their GitHub popularity.
Academia, by contrast, has been slower to adopt generative AI, in part because research cycles prioritize rigor and thoroughness. Research published at CHI 2023, for instance, relied on GPT-3, a model introduced in 2020 that is now considered outdated.

Scholarly discourse on generative AI frequently centers on its limitations (e.g., \citet{3592981.pdf}) and potential harms (e.g., \citet{2304.09991.pdf}). Many academic critiques offer normative guidance on what OpenAI’s LLMs should not be used for, even as their potential in real-world settings remains underexplored \citep{3544548.3581503.pdf}. Such value-laden narratives shape the direction of academic investigations \citep{1912.01172.pdf} and reflect a deep-seated skepticism toward OpenAI’s proprietary systems \citep{3608966.pdf}. The literature often overlooks LLMs’ potential for in-context reasoning (e.g., \citet{PARROTS}). While LLMs can hallucinate when tasked with zero-shot knowledge extraction \citep{3616863.pdf}, their strength lies in processing context-specific tasks effectively. Applications such as insight mining and knowledge discovery remain underexplored in this regard.

Our goal is not to extract absolute truths from text corpora but to distinguish meaningful signals from noise and to surface actionable insights that are ``good enough’’ for specific purposes. This orientation aligns with a pragmatic view of knowledge creation. As in machine learning, where hyperparameter tuning affects outcomes, effective prompt design for insight mining requires iterative experimentation and expertise in prompt engineering \citep{creativity,taxonomy}. This represents a departure from traditional research practices in HCI and other fields, where experiments typically yield singular outcomes. The polyvocality that LLMs afford adds value in this context \citep{3461778.3462100.pdf,a_polyvocal_and_contextualised_semantic_web.pdf,978-3-031-28244-7_18.pdf}: each output reflects a plausible interpretation, and small variations in prompts can reveal perspectives on textual data that would be difficult to achieve manually.
Our objective is to support experimentation from diverse perspectives in a cost- and time-efficient manner.
%
\subsection{Cost-efficient Data Exploration with LLMs}%
\label{sec:cost-efficiency}%
%
Research involves experimentation, and a corpus of scholarly papers can be examined from many perspectives. We might ask, for example: What are the current research topics in the HCI subfield of accessibility? Or: What methods do HCI researchers use? Answers to such questions are dispersed across papers in the HCI domain and often span multiple sub-communities. Repeatedly prompting LLMs with the entire document corpus to address such questions demands an agile and cost-efficient approach. Cost-efficient insight mining is particularly valuable when the goal is to analyze a text corpus from diverse perspectives. \citet{3544548.3580907.pdf}, for instance, showed how journalists might analyze a press release from multiple angles. Our study demonstrates how such an approach can scale from individual documents to entire text corpora.

The CHI conference proceedings served as an ideal case study. The 2023 proceedings alone encompass approximately 11,894 pages of content. To put this in perspective, if these papers were printed and laid end-to-end, they would stretch about 3.32 kilometers (2.06 miles). For a single researcher, reviewing this volume of literature to extract key HCI research challenges would require weeks or months of dedicated effort. Our application of LLMs reduced the time to approximately 17.5 hours on a standard cloud computing environment. Because the extraction ran as API calls, our team could attend to other activities in the meantime, illustrating a practical advantage of LLM-based workflows for academic research.

The cost-efficiency of our approach could be further improved by using free, open-source alternatives to OpenAI’s proprietary APIs. Options include Stable Beluga \citep{beluga}, Falcon \citep{Falcon}, T5 \citep{T5}, Pythia \citep{Pythia}, StableLM \citep{StableLM}, or LLaMA \citep{10000000_663429262362723_1696968207443577320_n.pdf,LLaMA}, along with their fine-tuned variants. OpenAI’s LLMs, however, remain notable for their performance \citep{gpt-4.pdf,2304.13712.pdf,AlpacaEval} while being relatively cost-efficient, and ChatGPT is accessible through straightforward APIs that require neither advanced machine learning expertise nor extensive local computing resources.

Although the analysis could be conducted entirely with the more capable but expensive GPT-4, cost-efficiency matters for flexible analysis of large text corpora. At the time of our study, GPT-3.5 completions were priced at US\$0.002 per 1,000 tokens, whereas GPT-4 completions (outputs) and prompts (inputs) were priced at \$0.03 and \$0.06 per 1,000 tokens, respectively. In our experiments, input prompts (which include the source text) accounted for the majority of the cost. Conducting all requests with GPT-4 would have increased the estimated total cost to approximately US\$512, compared to around US\$50 with the combination of GPT-3.5 and GPT-4.
%
%
\subsection{LLMs -- A Hammer for Every Nail?}%
\label{sec:reviewerbelief}%

While LLMs excel in many real-world tasks, they are not a universal solution for all problems \citep{IJHCI4}. Much of the skepticism surrounding LLMs stems from their misuse. A primary critique is that LLMs merely predict the next token and are inherently unreliable because of the noise and biases in their web-scraped training data \citep{2021.acl-long.330.pdf,1707.09457.pdf,2207.10245.pdf}. User trust is a prerequisite for AI adoption \citep{IJHCI3,10447318.2022.2095478}, yet LLMs exhibit a tendency toward sycophancy, that is, they generate answers that align with a user’s preferences even when those answers are suboptimal \citep{2212.09251.pdf,2310.13548.pdf}. As autoregressive models, they commit to their initial outputs, which may lead to ``hallucination snowballing,’’ where false claims compound over time \citep{2305.13534.pdf}. Techniques such as backtracking and pause tokens \citep{2306.05426.pdf,2310.02226.pdf} have been proposed to address these failure modes, but such methods were beyond the scope of our investigation.

We argue that hallucinations \citep{3571730.pdf,2305.13534.pdf} and sycophancy \citep{2310.13548.pdf,2212.09251.pdf} are less concerning when LLMs are used in a knowledge-intensive, context-grounded manner. In our task, correct usage involves providing the model with a clear and relevant context rather than posing standalone questions. Under these conditions, we found OpenAI’s LLMs capable of answering context-specific questions without hallucinating. Prompt engineering can further instruct the model to explicitly state when an answer is unknown, which reduces the risk of fabricated responses.

A more fundamental limitation of autoregressive models is their inability to generate genuinely new ideas. While they can retrieve and recombine information from their training data, their outputs remain bounded by that data. In qualitative research, depth often requires iterative reflection, what researchers describe as ``sleeping on the data.’’ This deliberate, time-intensive process allows researchers to learn from the experience of conducting the analysis itself, an aspect that is lost when tasks are delegated to LLMs. The contrast points to a potential misalignment between the rapid processing capabilities of LLMs and the traditional, immersive character of qualitative inquiry.

That said, the researcher’s personal learning journey is inherently tacit and difficult to share \citep{2635068.pdf}. LLMs, despite their internal opacity, allow quick dissemination of findings without interpretive translation by individual researchers. LLMs may also produce less biased outcomes, though this hypothesis requires further empirical validation. Finding the right balance between LLM efficiency and the depth of qualitative inquiry remains an important direction for future research.

\subsection{Implications for HCI Practice, Education, and Policymaking}%
\label{sec:hci-implications}%

As AI becomes increasingly integrated into work and research environments \citep{IJHCI1,IJHCI6}, the HCI community faces important questions about the ethical implications of deploying LLMs in research \citep{IJHCI5} and education \citep{IJHCI7}. Below, we consider implications for research practice, pedagogy, and policy.

\subsubsection{Implications for HCI practice}%

HCI research has a noticeable gap in large-scale investigations, largely because traditional qualitative methods are difficult to scale. Most studies involve only a limited number of participants \citep{small-n}, which is unsurprising given the costs and effort that studies with human subjects demand. Face-to-face research is expensive and labor-intensive, and the qualitative approach, frequently rooted in grounded theory \citep{glaserstrauss}, is constrained not by the availability of data but by the capacities of the human investigator. While grounded theory has supported in-depth work and small-N investigations hold undeniable value \citep{s13423-018-1451-8.pdf}, findings from such studies often face challenges in generalizability. Grounded theory has also been criticized as ill-suited for contemporary HCI studies that involve large datasets \citep{deterding2018.pdf}. Despite these limitations, there has been relatively little effort to scale research in HCI, even though doing so could broaden the scope and impact of findings.

Large language models may help address this gap by reducing many bottlenecks in qualitative research, including the labor-intensive and potentially biased nature of manual analysis. \citet{3544548.3580907.pdf} demonstrated that LLMs can analyze a single text document from multiple perspectives, and our study extends this idea to the scale of a large conference proceedings corpus. Conducting qualitative analyses from multiple perspectives manually is often impractical because of the inherent biases of researchers and the time investment required. With LLMs, researchers can analyze textual data from diverse perspectives while extending qualitative analysis beyond single documents, which represents a meaningful opportunity to scale HCI research.

In our work, we employed large language models with in-context learning \citep{2005.14165.pdf,2204.02329.pdf} to mine insights efficiently. We prioritized this approach over statistical topic modeling techniques because it allowed us to analyze textual data in line with specific objectives, in our case, extracting research challenges. Using LLMs, we scaled our analysis to encompass the entire CHI 2023 proceedings, which reduced issues such as human fatigue and the repetitive nature of qualitative coding.

We encourage the HCI community to explore new ways of integrating LLMs into research methodologies. A systematic approach to prompt design will be important for more streamlined and expansive (automated) knowledge accumulation and could deepen our understanding of user experiences and behaviors at a scale that manual methods cannot match. Beyond expanding the scope of investigations, LLMs may also enhance the depth and nuance of insights by surfacing patterns and trends that smaller-scale studies might overlook.%

Beyond academic research, the approach demonstrated in this study may also be relevant to industry practitioners. UX researchers and product teams regularly accumulate large volumes of qualitative data from user interviews, usability tests, and design reviews. Applying LLM-based extraction methods to such corpora could allow practitioners to systematically surface recurring pain points, unmet needs, and design opportunities that might otherwise remain buried in unstructured notes. While further validation would be needed to adapt our methodology to practitioner contexts, the underlying principle of scaling qualitative insight extraction appears transferable.

\subsubsection{Implications for HCI education}%

The research challenges identified in this study can inform HCI education by highlighting emerging areas of focus. They may shape curriculum development and pedagogical strategies that better prepare students for future challenges. Combined with the interactive visualization, this work may inspire new research directions in HCI and serve as a resource for understanding the field’s current state. The visualization offers a snapshot of prevailing topics in HCI, which makes it particularly valuable for early-career scholars and newcomers to the discipline.

In graduate seminars, for example, instructors could use the interactive visualization to orient new students to the breadth of HCI and to facilitate structured discussions about where open problems cluster. The dataset of extracted challenges could serve as a starting point for literature review exercises in which students identify gaps in existing knowledge and formulate their own research questions. More broadly, making such resources available to students may lower the barrier to developing a panoramic understanding of the field, a perspective that typically requires years of reading and conference attendance to acquire.

However, our approach also presents challenges, particularly if the methodology is applied without thoughtful reflection. Generative technologies such as language models may tempt some researchers to engage in superficial inquiries aimed at producing quick results \citep{3699689.pdf}. Addressing this issue will require targeted training and curriculum development in higher education \citep{10447318.2024.2352919}. Attitudes toward AI in education are evolving and may influence its adoption in academic programs \citep{10447318.2024.2365453,10447318.2024.2405786,10447318.2024.2401249}.

\subsubsection{Ethical and policy implications}%

Given the broad capabilities of large language models, it is essential to address their inherent biases and establish oversight to prevent misuse or misinterpretation of data. Collaboration between AI specialists and HCI researchers can support the development of tools tailored to the specific needs of HCI research and promote interdisciplinary approaches to knowledge creation.
The interactive visualization and intertopic distance map can further support interdisciplinary collaboration by identifying overlapping interests and complementary expertise across disciplines. This could inspire the establishment of research centers dedicated to addressing ethical AI challenges from multiple angles.

The findings of our study also have implications for policy-making, particularly in promoting ethical AI use in academia. By systematically identifying and visualizing current research challenges and their clustered topics, this approach highlights important knowledge gaps and can inform decision-making for academic institutions, funding agencies, and regulatory organizations.

One implication is the potential to influence funding priorities. The dataset of research challenges provides a strategic resource for funding agencies to allocate resources effectively. Challenges related to ethical AI development, such as bias mitigation, transparency, and equitable access, can be prioritized. Additionally, clusters linked to the UN Sustainable Development Goals (SDGs) offer a framework for policymakers to direct funding toward areas with high societal relevance.

Another area of impact is guiding institutional policies on AI adoption. Universities and research organizations can draw on findings from this study to craft guidelines for responsible AI use in research. Such policies could emphasize fairness, reproducibility, and inclusivity in line with evolving ethical standards.

Finally, this study highlights the importance of transparency in addressing challenges within academic research. By making the dataset and visualization publicly available, we demonstrate a commitment to openness and accountability. Policymakers can use this transparency to promote similar practices across academic and industrial AI research and to encourage a culture of ethical responsibility and open inquiry.

Beyond HCI, the methodology we present may transfer to other disciplines in which authors routinely discuss open problems in published literature. For example, applying the two-step extraction approach to medical informatics proceedings (such as AMIA or MEDINFO) could surface unresolved questions in clinical decision support, electronic health records, or patient-facing technologies. The approach could likewise be adapted to identify pedagogical challenges from educational research venues (such as SIGCSE or LAK), or to extract policy gaps from legal and regulatory corpora. These cross-disciplinary applications would require only modest changes to prompt design. The core method, which provides source text as context and instructs the model to extract domain-specific statements, remains applicable regardless of the target field.

\subsection{Limitations}%
\label{sec:limitations}%
In this paper, we investigated the knowledge-intensive task of extracting research challenges from scholarly articles. While our findings are grounded in HCI research, we do not claim generalizability to other tasks or text corpora. We believe, however, that the methodology has potential applicability across various academic disciplines, given the versatility of LLMs in insight mining and analysis. LLMs can facilitate qualitative analysis from diverse perspectives and reduce the cognitive workload associated with such tasks \citep{IJHCI2}.

Using LLMs for insight mining poses challenges, including ensuring the relevance and accuracy of extracted information and mitigating inherent model biases. These challenges point to the need for careful prompt design and validation. A key limitation of our approach is the requirement for upfront knowledge about what questions can be answered from the text corpus. In our case, extracting research challenges was feasible because most CHI papers mentioned at least one challenge. Conversely, attempting to identify factual statements not present in the corpus would likely lead to hallucinations \citep{2301.00303.pdf}. Prompt engineering could address this issue to some extent by instructing the model to acknowledge when an answer is unknown, and future work could explore improved prompt designs for such scenarios.

Our experiments also revealed that results are sensitive to prompts, which highlights the importance of prompt design (or prompt engineering) as an emerging research area. Norms for prompt design remain underdeveloped, however, and current practices often lack standardization. From our work, we observed that once a well-designed prompt was identified, results were stable and reliable with minimal deviations (cf. Section \ref{sec:noise}). Beyond prompt design, consistent querying of the LLM API is important for reliable outputs. For some papers, we queried the API in smaller batches to avoid exceeding GPT-3.5’s context window limit. Future research could investigate how batch size affects output consistency and reliability.

Another concern is the opacity of OpenAI’s latest LLMs. These models may encode latent opinions \citep{3544548.3581196.pdf} and cultural values \citep{2203.07785.pdf} and may reproduce stereotypes and biases \citep{2005.14165.pdf,Bommasani2021FoundationModels}. While we cannot fully exclude the presence of such biases in our data, a qualitative evaluation of a sizable sample did not reveal significant traces of bias. We encourage the research community to scrutinize the extracted research challenges, which are openly available for exploration and download.

Several additional methodological constraints merit acknowledgment. First, our analysis draws on a single corpus, the CHI 2023 proceedings, which means the extracted challenges reflect one venue and one year. Research challenges discussed at CHI may differ from those at other HCI venues such as CSCW, UIST, or DIS, and a multi-venue corpus would yield a broader picture of the field. Second, the qualitative evaluation of extracted challenges relied on a single annotator, which limits the assessment of inter-rater reliability. While the quantitative metrics (precision, recall) provide complementary evidence, future evaluations would benefit from involving multiple independent annotators. Third, the findings represent a temporal snapshot: the challenges captured here reflect the state of HCI research as of early 2023, and both the field’s priorities and LLM capabilities have continued to evolve since.
%
\subsection{Future Work}%
\label{sec:future-work}%

Language models offer a flexible means for mining insights from data and represent an alternative to traditional natural language processing techniques. However, barriers remain that hinder their widespread adoption.

One barrier is social. Many researchers have spent decades developing traditional approaches for mining insights, and there is understandable skepticism about using LLMs for this purpose. Future work could focus on increasing the social acceptability of such methods.

Another barrier is technical. While in-context learning has significantly improved LLM performance, issues such as hallucinations and non-factual statements persist. Current technologies are unlikely to achieve perfect accuracy, and a shift in perspective is needed. Just as 80\% accuracy is often deemed sufficient for publication in machine learning, we argue that similar thresholds in insight mining can yield valuable and useful results. Nonetheless, future work should continue to improve accuracy and reduce the potential for hallucinations.

Another area for exploration is addressing LLMs’ tendencies to generate biased or irrelevant responses. Mitigation strategies include iterative prompt refinement and cross-validation with domain experts. Feedback loops, such as incorporating expert validation of extracted challenges, could adopt a human-in-the-loop approach to enhance the accuracy, relevance, and reliability of results. Such mechanisms would also support a more collaborative and human-centered approach to data analysis.

Beyond these general considerations, several concrete directions emerge from the work presented in this article. First, applying our extraction pipeline to proceedings from multiple years would allow longitudinal tracking of how research challenges evolve, which topics gain or lose prominence, and whether the community's stated challenges align with subsequent research output. Such temporal analysis could reveal whether certain challenges persist across years or whether they are resolved, reframed, or superseded by new concerns. Second, the rapid development of language models since our data collection (including GPT-4o, Claude, Gemini, and open-source alternatives such as LLaMA and Mistral) raises the question of how model choice affects extraction quality. Systematic comparisons across models would help identify which aspects of challenge extraction are model-dependent and which remain stable, and would inform best practices for replication and extension of this work. Third, combining LLM-based extraction with structured expert validation workshops could strengthen both the credibility and the utility of the results. In such a design, domain experts would review and annotate LLM-extracted challenges, to produce curated datasets that serve dual purposes: as validated research outputs and as training signal for improving future extraction pipelines.

\section{Conclusion}%
\label{sec:conclusion}%
In this paper, we presented an approach that uses large language models to identify key research challenges in the field of HCI. Specifically, we applied ChatGPT (GPT-3.5 and GPT-4) to extract statements from the 2023 CHI proceedings.
The identified challenges are visualized in an interactive map that provides an overview of the current state of HCI research.
%
This study demonstrates the potential of LLMs to mine insights from extensive text corpora and offers a methodology that supports both HCI research and broader academic inquiry.
Our approach provides a way for researchers to better understand the focus of a research community and track how it evolves over time. The proposed two-step process proved to be both efficient and cost-effective.

While LLMs offer exciting opportunities for analyzing new data sources and extracting insights at scale, our approach is not intended to replace qualitative analysis. We used BERTopic, a method for qualitative and iterative topic exploration, to analyze the research challenges identified. Rather than supplanting qualitative methods, our approach enhances the ability to explore data sources that are infeasible to analyze manually.

The field of LLMs is evolving rapidly.
Since completing our study, fine-tuned variants of open-source models have become available via APIs from providers such as Amazon Bedrock, Microsoft Azure, and Huggingface.
Given the trend toward closed-source generative models, API-based research is likely to become the norm \citep{2304.13712.pdf}.
These APIs offer a flexible and scalable approach to data analysis and may make LLM-based scientific discovery a standard practice.
In our study, we applied LLMs to analyze a corpus of scholarly documents from the CHI 2023 proceedings with the goal of discovering research challenges. Through iterative prompt design and careful evaluation, we found that combining ChatGPT's GPT-3.5 and GPT-4 models provides an effective means of qualitatively analyzing a text corpus from a specific perspective. This method of insight mining can be repeated for various perspectives, and the two-step process is cost-efficient.

Evaluating LLMs on practical tasks, such as ours, is important to inform researchers and practitioners about the LLMs' true capabilities. Misjudging the performance of LLMs can hinder their adoption in research. We encourage researchers to apply our approach to derive insights from textual data at scale.

\section*{Disclosure Statement}
In accordance with Taylor \& Francis policy, the authors disclose the following use of generative AI in the preparation of this manuscript. Claude Code (Anthropic, Claude Opus 4, 2025) was used to assist with proof-reading and copy-editing of the manuscript text. The authors reviewed and edited all AI-assisted output and take full responsibility for the content of this publication.

\section*{Competing Interests}
The authors declare no competing interests.

\section*{Funding}
This research received no specific grant from any funding agency.

\ifshowrewrite\color{black}\fi
\bibliographystyle{apacite}%
\bibliography{paper}%


\clearpage
\appendix
\section{HCI Research Topics}%
\label{appendix-topics}%
\input{APPENDIX-TOPICS}

\section{Prompts}%
\label{sample-prompts}%

\subsection{Step 1}%
Forget all the above. You are InfoMinerGPT, a language model trained to extract challenges for
researchers in the field of HCI from academic papers. You output a list of challenges, one per
line, without explanations. If there are no challenges mentioned in the text, you output ’None’.
Here is the academic paper:{\textbackslash}n{\textbackslash}n

\subsection{Step 2}%
Below is a JSON list of challenges extracted from an academic paper. Please remove any dupli-
cates. Then keep only the 5 most important challenges for researchers in the field of HCI. Select
research challenges based on their relevance to current HCI trends, potential for future impact,
and novelty. You output a list of challenges, one per line, without explanations. Here are the
challenges:{\textbackslash}n{\textbackslash}n


\section*{Biographies}
\vspace{1em}

\noindent
\textbf{Jonas Oppenlaender}
Jonas Oppenlaender is a Postdoctoral Researcher at the University of Oulu. His research focuses on generative AI, prompt
engineering and human-centred AI. He teaches the master-level course AI Engineering, where students design and build
multi-agent systems and gain hands-on experience with LLM-based agents, orchestration, tool use and retrieval-augmented generation (RAG). 
He holds a Doctor of Science (Tech.) in Computer Science (with distinction) and a Microsoft Certified Azure AI Engineer Associate certification. Prior to
academia, he worked as a full-stack web developer and in roles spanning business analysis and IT service design.
\vspace{2em}

\noindent
\textbf{Joonas Hämäläinen}
is a postdoctoral researcher at the University of Jyväskylä. He received the B.S. degree in physics, the M.S. degree in applied physics, and the Ph.D. degree in mathematical information technology from University of Jyväskylä, Jyväskylä, Finland, in 2012, 2013, and 2018. His main research interests include data mining and machine learning.

\end{document}

%% file: FIG-PROCESS.tex


\begin{figure}[!htb]%
    \centering%
    \includegraphics[width=\linewidth]{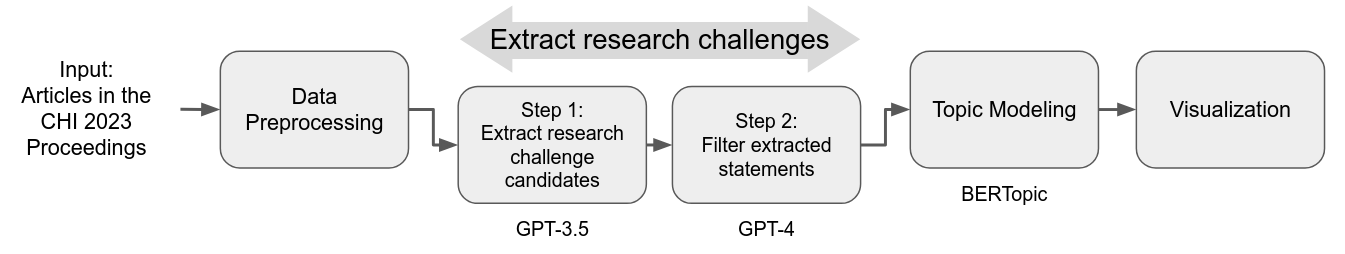}
    \caption{Overview of the process for extracting research challenges from the corpus of scholarly articles. The pipeline consists of five stages: (1)~data preprocessing to clean and normalize the extracted text, (2)~broad extraction of candidate challenges with GPT-3.5, (3)~filtering and ranking by GPT-4 based on relevance, impact, and novelty, (4)~topic clustering with BERTopic, and (5)~interactive UMAP visualization of the resulting challenge landscape.}%
    \label{fig:process}
\end{figure}%

%% file: FIG-CHALLENGES.tex
\begin{figure*}[thb]%
\centering%
\includegraphics[width=\textwidth, trim={1.1cm 1.4cm .3cm 1.15cm},clip]{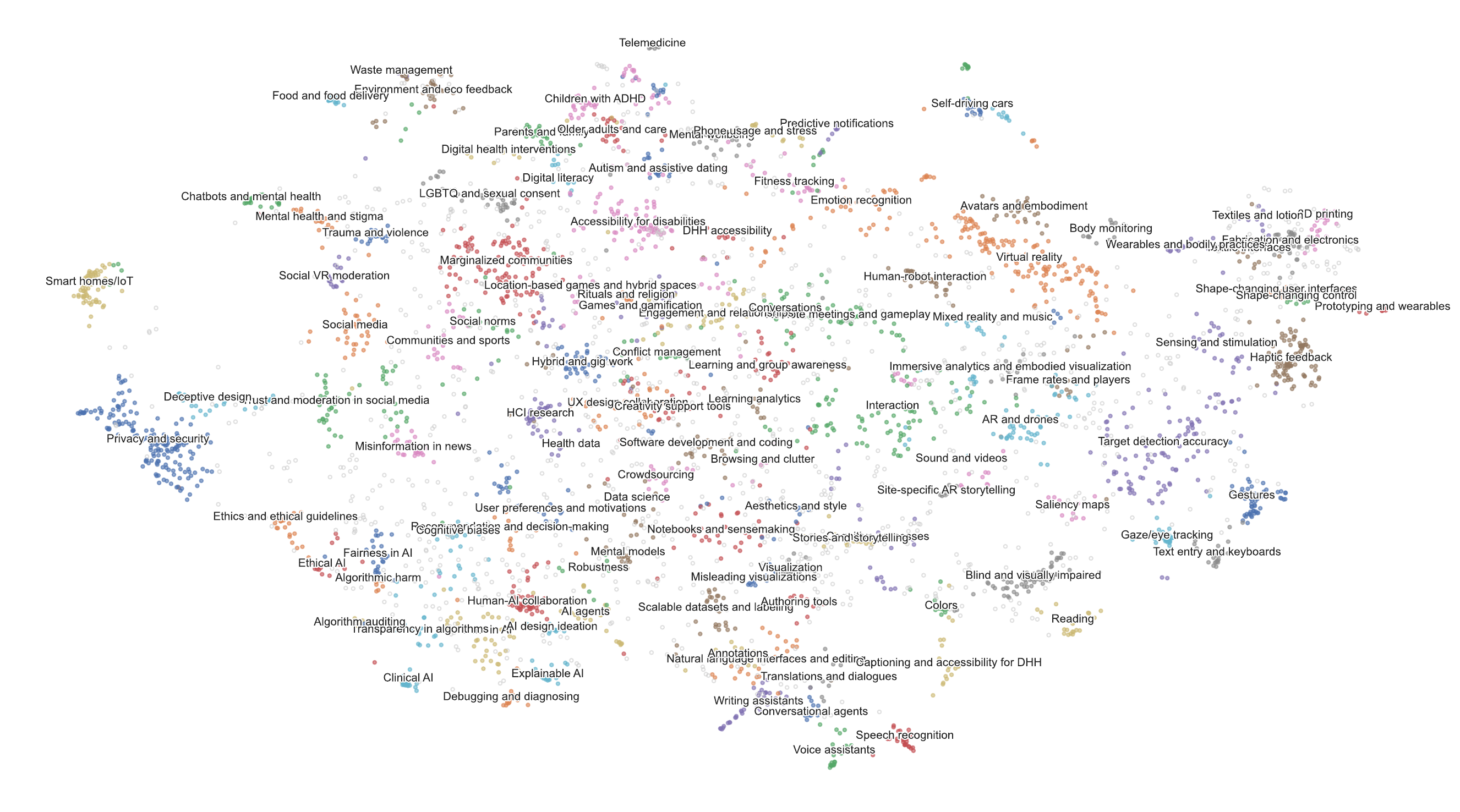}%
\caption{Research challenges extracted from CHI 2023 proceedings using ChatGPT, with topic annotations by BERTtopic \citep{grootendorst2022bertopic}.
To create the plot, the challenges were first converted into embeddings 
and then projected into two-dimensional space with UMAP \citep{UMAP}.
Each filled circle represents a research challenge, with colors indicating different topic clusters identified by BERTopic. Gray, unfilled circles represent challenges that did not group into any of the identified clusters. 
An interactive visualization is available at \href{https://hci-research-challenges.github.io/}{https://hci-research-challenges.github.io}.%
}%
\label{fig:viz}%
\end{figure*}%

%% file: FIG-GC-COMPARISON.tex
\begin{figure}[p] 
  \centering
  \rotatebox{90}{
    \begin{minipage}{\textheight}
      \centering
      \includegraphics[width=0.9\textheight]{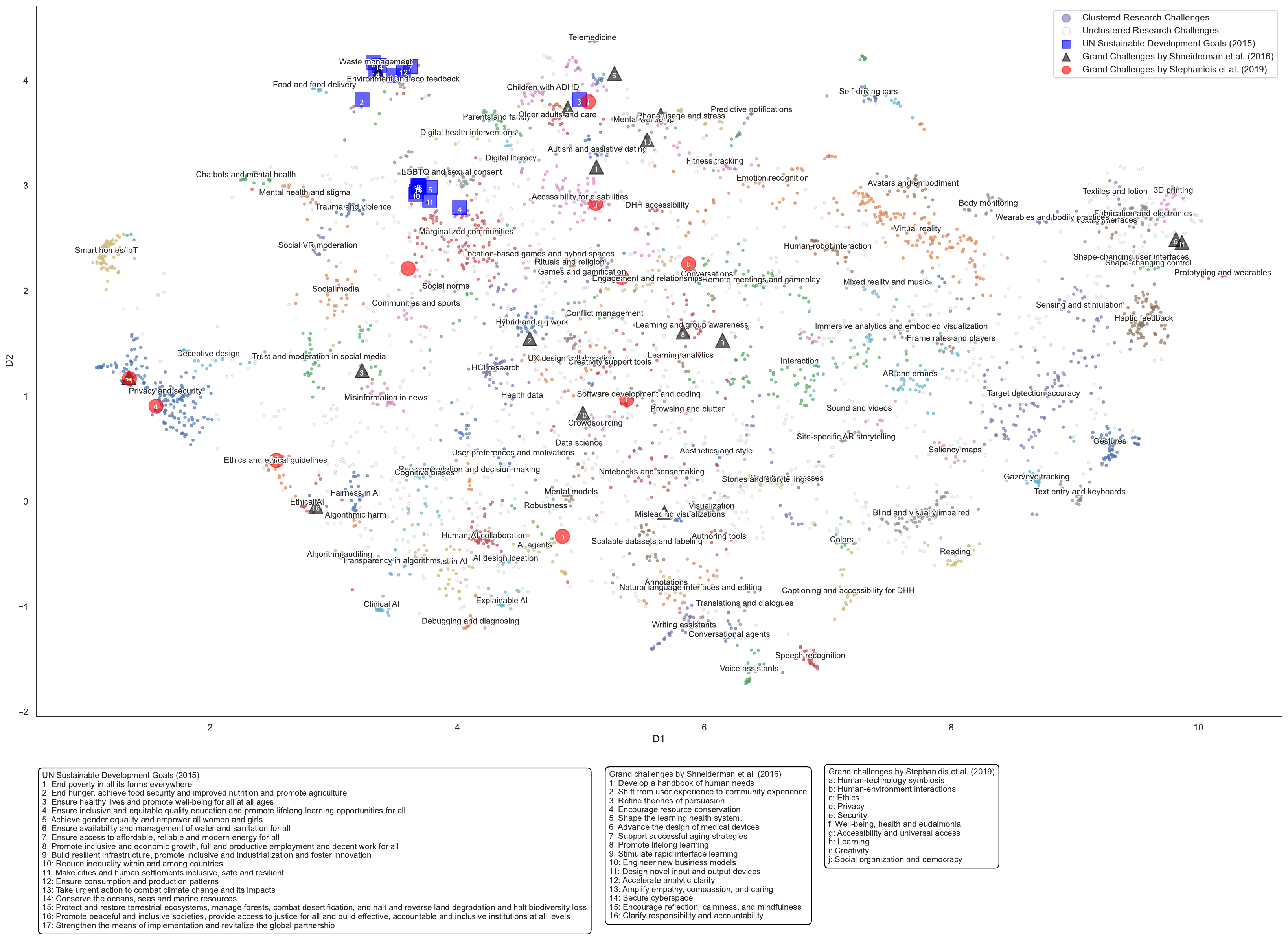}
      \caption{Comparison of extracted research challenges with HCI's grand challenges and the United Nations' Sustainable Development Goals (SDGs). Each small dot represents one of the 4,392 research challenges extracted from the CHI 2023 proceedings, color-coded by topic cluster. Grand challenges and SDGs are plotted as labeled markers in the same embedding space. Proximity between a grand challenge marker and a cluster of research challenges indicates thematic alignment. The SDG labels were modified to reduce the influence of the shared term ``sustainable'' on the embedding positions (see Section~\ref{sec:results}).}
      \label{fig:gccomparison}
    \end{minipage}
  }
\end{figure}

%% file: TAB-METRICS.tex
\begin{table}[!htb]%
\caption{Insight mining performance, as evaluated by canonical metrics from the field of NLP.}%
\label{tab:metrics}%
\centering
\footnotesize
\begin{tabular}{lccccc}%
\toprule%
     & EM & ED & WER & BLEU & ROUGE-1 \\
\midrule%
    Step 1\textsuperscript{†} &
        0.0878 &
        56.8213 &
        0.9703 &
        48.53 &
        0.6822 \\
    Step 2\textsuperscript{‡} & 0.4571 & 2.0260 & 0.0591 & 35.70 &  0.9713 \\
\midrule%
      & BLANC & BERTScore  & METEOR & 
      BLEURT-20 & ROUGE-L \\
\midrule%
    Step 1\textsuperscript{†} &
        0.4345 &
        0.9361 & 
        0.6058 &
        0.6755 & 
        0.6554 \\
    Step 2\textsuperscript{‡} & 
    0.6840 &
       0.9629  %
    & 0.9812 &
      0.9459 &
    0.9712  \\
\bottomrule%
\multicolumn{6}{l}{\footnotesize\textsuperscript{†}%
Best-matching statements between statements extracted with GPT-3.5 and source sentences.
}\\
\multicolumn{6}{l}{\footnotesize\textsuperscript{‡}%
Best-matching statements between statements extracted by GPT-4 and GPT-3.5.
}
\end{tabular}%
\end{table}%

%% file: TAB-EXAMPLE.tex
\begin{table*}[t]%
\caption{Excerpt of information extracted from one CHI 2023 paper~\cite{3544548.3580902}.
    Research challenges selected by GPT-4 (left)
    from the research challenges extracted by ChatGPT, compared to the source text (right).
    The sentences from the source text in this table were selected based on embedding cosine similarity to the ChatGPT sentence embeddings.%
}%
\centering
\label{tab:example}%
\begin{tabularx}{\textwidth}{
    >{\hsize=.33\hsize}X
    >{\hsize=.65\hsize}X
}%
    \toprule
    \small
    GPT-4 challenge &
    \small
    Most similar sentence from source text
\\
    \midrule

\footnotesize
reimagining relationships with technology through self-discovery
&
\footnotesize
rather than use self-tracking devices to know or change ourselves, our process deploys other-tracking tracking devices (those designed for technology firms to collect our data) to reimagine our relationships with technology
\\
    \midrule

\footnotesize
appropriation and ownership of personal data
&
\footnotesize
through a reworking of personal data, the speculation emphasizes the work of adopting something for ones own use without the explicit permission of the entities that control that data
\\
    \midrule

\footnotesize
power dynamics embedded in corporately controlled data
&
\footnotesize
first, it orients analysis toward a consideration of the power dynamics embedded in corporately controlled data
\\
    \midrule

\footnotesize
emotional labor and affective positioning in autospeculation
&
\footnotesize
for kinnee, each encounter with the data requires a form of emotional labor that brings about strong physical and affective responses
\\
    \midrule

\footnotesize
polyvocality and the need to elevate diverse voices in autospeculation
&
\footnotesize
in this orientation, our analysis prompts a few important questions for hci: whose voices should autospeculation elevate?
\\

\bottomrule
\end{tabularx}%
\end{table*}%

%% file: APPENDIX-TOPICS.tex
\begin{table}[!htb]%
\caption{
Research topic clusters with the number of respective research challenges.}%
\label{tab:challenges}%
\footnotesize
\begin{tabularx}{\textwidth}{
    Xl@{\hspace{25pt}}
    Xl@{\hspace{25pt}}
    Xl
}%
\toprule%
    Topic & \# & Topic & \# & Topic & \# \\
\midrule%
    1. Privacy and security & 174 &
        39. AI agents & 28 &     
            77. Sound and videos & 16 \\
    2. Virtual reality & 139 &       
        40. Deceptive design & 28 &  
            78. Body monitoring & 16 \\
    3. Interaction & 107 &  
        41. Self-driving cars & 28 & 
            79. Phone usage and stress & 15 \\
    4. Marginalized communities & 106 & 
        42. UX design collaboration & 28 & 
            80. Transparency in algorithms & 15 \\
    5. Target detection accuracy & 103 &
        43. Chatbots and mental health & 26 &  
            81. \raggedright User preferences and motivations & 15 \\
    6. Haptic feedback & 95 &  
        44. \raggedright Learning and group awareness & 26 & 
            82. \raggedright Recommendation and decision-making & 15 \\
    7. Accessibility for disabilities & 77 & 
        45. Cognitive processes & 26 & 
            83. Conversations & 14 \\
    8. Blind and visually impaired & 70 &  
        46. \raggedright Scalable datasets and labeling & 26 & 
            84. DHH accessibility & 14 \\
    9. Smart homes/IoT & 67 & 
        47. Fitness tracking & 25 & 
            85. \raggedright Wearables and bodily practices & 14 \\
    10. AR and drones & 61 & 
        48. Mental wellbeing & 25 & 
            86. Learning analytics & 13 \\
    11. Gestures & 57 & 
        49. Reading & 24 & 
            87. Crowdsourcing & 13 \\
    12. Emotion recognition & 51 &  
        50. Food and food delivery & 24 &  
            88. Telemedicine & 13 \\
    13. \raggedright Trust and moderation in social media & 47 & 
        51. Fairness in AI & 24 &  
            89. Annotations & 13 \\
    14. Human-AI collaboration & 46 & 
        52. Natural language interfaces and editing & 24 & 
            90. Cognitive biases & 13 \\
    15. HCI research & 45 & 
        53. Voice assistants & 23 &  
            91. Misleading visualizations & 13 \\
    16. \raggedright Environment and eco feedback & 45 &  
        54. Older adults and care & 23 & 
            92. Rituals and religion & 13 \\
    17. Children with ADHD & 44 & 
        55. Sensing and stimulation & 22 & 
            93. Colors & xx \\
    18. LGBTQ and sexual consent & 40 &
        56. \raggedright Software development and coding & 22 & 
            94. Prototyping and wearables & 12 \\
    19. Trust in AI & 40 & 
        57. Communities and sports & 22 & 
            95. Waste management & 12 \\
    20. Clinical AI & 37 & 
        58. Text entry and keyboards & 21 &  
            96. Frame rates and players & 12 \\
    21. Hybrid and gig work & 37 & 
        59. Captioning and accessibility for DHH & 21 & 
            97. Browsing and clutter & 11 \\
    22. Social media & 36 &  
        60. Mixed reality and music & 21 & 
            98. Translations and dialogues & 11 \\
    23. Remote meetings and gameplay & 35 & 
        61. Trauma and violence & 21 & 
            99. Algorithm auditing & 11 \\
    24. Speech recognition & 35 &  
        62. Mental health and stigma & 20 &  
            100. Digital literacy & 11 \\
    25. Writing assistants & 35 &  
        63. Social norms & 20 & 
            101. Aesthetics and style & 11 \\
    26. Human-robot interaction & 35 & 
        64. Creativity support tools & 20 & 
            102. \raggedright Shape-changing user interfaces & 11 \\
    27. 3D printing & 35 &  
        65. Social VR moderation & 20 &  
            103. Shape-changing control & 11 \\
    28. Fabrication and electronics & 34 &  
        66. Mental models & 19 & 
            104. Authoring tools & 10 \\
    29. \raggedright Engagement and relationships & 33 & 
        67. \raggedright Location-based games and hybrid spaces & 18 & 
            105. Textiles and lotion & 10 \\
    30. Gaze/eye tracking & 32 & 
        68. \raggedright Immersive analytics and embodied visualization & 18 & 
            106. Data science & 10 \\
    31. Health data & 32 & 
        69. Stories and storytelling & 18 &  
            107. Saliency maps & 10 \\
    32. Ethics and ethical guidelines & 32 &  
        70. Explainable AI & 17 &  
            108. Site-specific AR storytelling & 10 \\
    33. Parents and family & 32 &  
        71. Autism and assistive dating & 17 &  
            109. Digital health interventions & 10 \\
    34. Notebooks and sensemaking & 31 &  
        72. Debugging and diagnosing & 17 & 
            110. AI design ideation & 10 \\
    35. Games and gamification & 30 & 
        73. Conflict management & 17 & 
            111. Conversational agents & 10 \\
    36. Avatars and embodiment & 30 &  
        74. Ethical AI & 17 & 
            112. Algorithmic harm & 10 \\
    37. Misinformation in news & 29 & 
        75. Predictive notifications & 16 & 
            113. Robustness & 10 \\
    38. Visualization & 29 &  
        76. Textile interfaces & 16 &  
             &  \\
\bottomrule%
\end{tabularx}%
\end{table}%